\documentclass[twocolumn]{article}
\usepackage[utf8]{inputenc}
\usepackage[top=2cm, bottom=3cm, left=2cm, right=2cm]{geometry}
\usepackage[T1]{fontenc}
\usepackage{authblk}
\usepackage{amsfonts}
\usepackage{amsmath}
\usepackage{amssymb}
\usepackage{xfrac}
\usepackage{graphicx}
\usepackage{float}
\usepackage{dblfloatfix}
\usepackage{placeins}
\usepackage{booktabs}
\usepackage{array}
\usepackage[numbers,sort&compress]{natbib}
\usepackage{comment}
\usepackage[skip=2pt,font=footnotesize]{caption}
\usepackage{titlesec}
\usepackage[flushleft]{threeparttable}
\usepackage{color}
\usepackage{xcolor}
\usepackage[normalem]{ulem}
\usepackage{tablefootnote}
\usepackage{caption}
\usepackage{subcaption}
\usepackage[colorlinks=true,urlcolor=blue,linkcolor=black,citecolor=blue]{hyperref}

\newcolumntype{L}[1]{>{\raggedright\arraybackslash}m{#1}}
\newcolumntype{M}[1]{>{\centering\arraybackslash}m{#1}}
\newcolumntype{R}[1]{>{\raggedleft\arraybackslash}m{#1}}

\titleformat*{\section}{\normalsize\bfseries\rmfamily}
\titleformat*{\subsection}{\normalsize\bfseries\rmfamily}
\titleformat*{\subsubsection}{\normalsize\bfseries\rmfamily}

\graphicspath{{./Figures/}}
\DeclareGraphicsExtensions{.jpg,.pdf,.pdf}
\title{\textbf{Integrated full pulse modeling for pellet injection in tokamaks: HPI2 model improvement and validation in WEST.}}
\author[1,2,*]{A. Panera Alvarez}
\author[3]{F. Koechl}
\author[4]{J. Artaud}
\author[4]{E. Geulin}
\author[4]{B. P\'egouri\'e}
\author[4]{E. Vergnaud}
\author[4]{C. Bourdelle}
\author[1]{S. Wiesen}

\author[a]{the WEST Team}
\affil[1]{DIFFER -- Dutch Institute for Fundamental Energy Research, 5612 AJ Eindhoven, the Netherlands}
\affil[2]{Eindhoven University of Technology, 5612 AZ Eindhoven, Netherlands}
\affil[3]{ITER Organization, Route de Vinon-sur-Verdon, 13067 St Paul Lez Durance Cedex, France}
\affil[4]{CEA Cadarache, IRFM, F-13108 Saint-Paul-lez-Durance, France}

\affil[a]{See \href{http://west.cea.fr/WESTteam}{http://west.cea.fr/WESTteam}}
\affil[*]{E-mail: \href{mailto:a.paneraalvarez@differ.nl}{a.paneraalvarez@differ.nl}}

\begin{document}

\twocolumn[
  \begin{@twocolumnfalse}
    \date{}
    \maketitle
    \begin{abstract}
    Reliable modeling and control of core density is essential for reactor-relevant magnetic confinement fusion operation, motivating cryogenic pellet injection as a primary fueling actuator and the need for predictive pellet source models in integrated modeling. Here we present an upgrade of the physics-based pellet code HPI2 in which the plasmoid release spatial step is determined self-consistently from ablation physics, $dx_{var}=v_{\mathrm{pel}}\,t_{\mathrm{exit}}$ (optionally rescaled to trade accuracy for computational cost), removing an ad-hoc discretization parameter and improving numerical robustness across injection conditions. The upgraded model is first validated in stand-alone against a high-field-side pellet-fueled, ohmic, WEST discharge (\#58656) by comparing synthetic and measured interferometry line-integrated density increments, obtaining a mean error of $\sim 10\%$. We then perform full-radius, time-dependent integrated modeling validation by coupling the new HPI2 within the High Fidelity Pulse Simulator (HFPS) workflow (JINTRAC/IMAS), combining JETTO with SANCO for the impurity/radiation evolution and TGLF-SAT2 for the turbulent transport. The coupled simulations reproduce the main density rise and relaxation after pellet injection and the associated electron-temperature transient, while taking into account the strong influence of tungsten radiation in WEST, supporting the consistency of HPI2 as a predictive pellet particle source in integrated modeling frameworks. Ultimately, this validation study supports the use of pellet modeling tools in integrated modeling studies for larger devices such as ITER.
    \end{abstract}
  \end{@twocolumnfalse}
  ]

\section{Introduction} \label{sec:pellet_fuelling}
Achieving and sustaining reactor-relevant fusion performance requires reliable control of the core particle inventory. Conventional fueling by peripheral gas puffing becomes increasingly ineffective as device size increases because the neutral ionization length remains roughly unchanged, whereas the distance between the gas-puff valve and the confined plasma becomes larger \cite{Mordijck_2020}, consequently increasing plasma neutral opacity. Neutral beam injection (NBI) can provide core fueling but intrinsically couples particle and energy sources and is therefore constrained by scenario requirements, plus its contribution depends on plasma volume and NBI power. Cryogenic pellet injection, proposed as a primary fueling actuator for ITER/EU-DEMO class devices \cite{Doyle_2007,Lang02012024}, offers rapid density control and deeper particle deposition by delivering a macroscopic amount of frozen hydrogenic material across flux surfaces. For ITER baseline H-mode conditions, integrated scenario studies indicate that pellets with typical masses of order $10^{21}$ atoms injected at a few hertz can provide the particle source required to sustain target density \cite{Marin_2021,Garzotti_2019}. Pellet fueling is thus expected to be central to density control and burn optimization in next-step tokamaks.

\begin{figure}
    \centering
    \includegraphics[width=\linewidth]{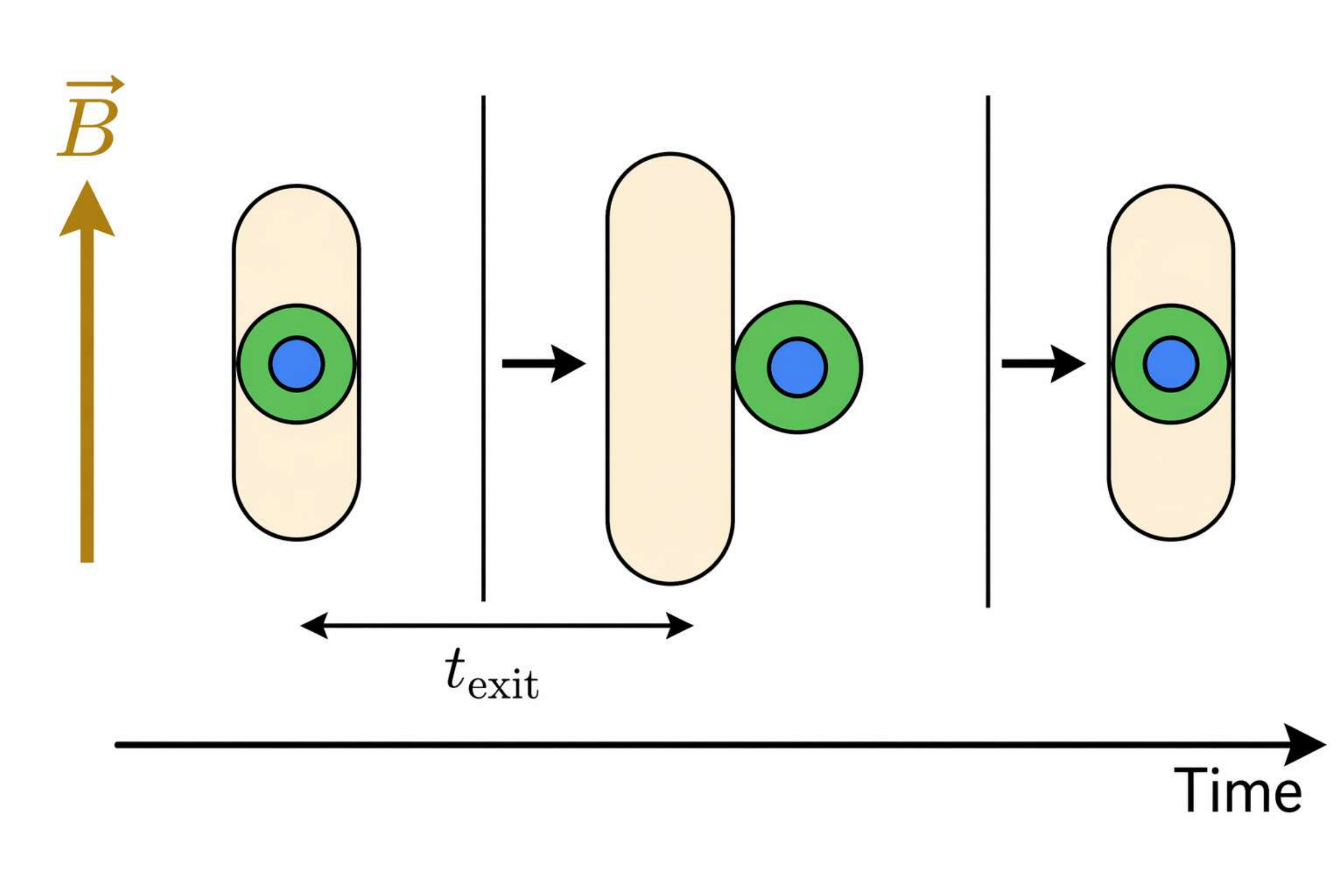}
    \caption{Scheme on the cyclic pellet ablation process in a tokamak plasma, showing $t_{exit}$ as the time that this process takes to repeat. The blue circle denotes the pellet, the green circle the neutral cloud, and the light-brown structure the plasmoid (ionized cloud). The horizontal axis represents time, while the vertical black lines separate successive stages of the ablation process.}
    \label{fig:t_exit}
\end{figure}

The pellet-plasma interaction is governed by a sequence of coupled processes occurring from microseconds to millisecond timescales. Upon entering the confined plasma, the incident heat flux from the hot background plasma drives rapid surface sublimation and ablation, producing a dense, cold cloud around the pellet providing strong self-shielding, acting as a first barrier to the incoming heat flux (often compared to a Leidenfrost-like effect). As collisions continue, the outer part of this cloud becomes progressively ionized, producing a plasma cloud (or plasmoid) that is guided along magnetic field lines and separates from the pellet, while fresh neutrals are continuously generated by ongoing ablation. This results in a quasi-cyclic ablation-ionization-ejection sequence as the pellet propagates inward (schematically illustrated in figure~\ref{fig:t_exit}), which controls the instantaneous ablation rate and ultimately the penetration depth. Several shielding mechanisms contribute to reducing the effective heat flux to the pellet: the dominant one is Coulomb stopping in the cloud (heat flux absorbed by collisions with plasma cloud and neutral cloud), with additional contributions e.g. from an electrostatic sheath at the cloud interface (principally repelling electrons). Moreover, in the presence of auxiliary heating, suprathermal electrons can modify the ablation dynamics and lead to earlier ablation, thereby limiting pellet penetration. For a more complete review see  \cite{Pegourie_2007}.

Ablated material is released as a sequence of dense, cold plasmoids along the pellet trajectory. These structures expand rapidly along the magnetic field until pressure homogenization with the surrounding plasma is achieved. During this phase, the inhomogeneous magnetic field induces a cross-field drift of the deposited material down $\nabla B$: in tokamaks, the opposite vertical motion of cloud electrons and ions produces a charge separation, described by a current $j_{\nabla B}$ which is only partially compensated by a polarization current $j_{\mathrm{pol}}$. The resultant electric field gives rise to an $E\times B$ drift of the plasmoid center-of-mass towards the Low Field Side (LFS) direction, i.e. outwards from the torus center, which correspond to the direction of the gradient of the magnetic field. This drift can substantially shift the effective deposition location relative to the geometric injection path; in particular in tokamaks it tends to expel material for low-field-side (LFS) injection while favoring penetration for high-field-side (HFS) injection. The drift process is further influenced by Alfv\'en wave emission and the formation of external and internal Pfirsch-Schlueter currents along magnetic flux tubes, which contribute to the reduction of the drift velocity, impacting the matter deposition profile depending on the specific magnetic configuration of the device. Particularly, the drift displacement depends on the safety factor profile and its resonant surfaces as demonstrated in \cite{Sakamoto_2013}. A quantitative description of parallel expansion, $\nabla B$-driven drift, and drift braking mechanisms is essential for predictive simulations of pellet-fueled discharges and for the interpretation of dedicated experiments.

A state-of-the-art model used in integrated modeling for pellet fueling is HPI2 (Hydrogen Pellet Injection) \cite{Pegourie_2007_HPI2,koechl2012modelling}. This model has been applied and validated in different studies: stand-alone and validation in different devices such as JET \cite{Koechl2010EPS_pelletJET,Koechl2007EPS_pelletDriftJET,Frigione2010JET_pelletDeposition,Marin_2021,valovivc2019control}, Tore Supra \cite{Commaux_2010,Sakamoto_2013,Klaywittaphat2011EPS_pelletToreSupra}, MAST \cite{Garzotti_2010}, EAST \cite{Huang_2024,Zhang_2024}, and non-axysimmetric devices \cite{panadero2018experimental,pegourie2024structure, panadero2023using,Matsuyama_2012,Baldzuhn_2019}; in future devices such as DTT, ITER, EU DEMO, and CFEDR \cite{LANG2020111591,Baiocchi_2023,Na_2019,Garzotti_2019,ZHANG_2025}; in integrated modeling frameworks \cite{Marin_2021,valovivc2019control}; or for control-oriented applications \cite{orrico2025predictive}. HPI2 summarizes the described phenomena, and it is improved and used in the present study. In the section \ref{sec:hpi2_up} the model is described further in detail.

Because pellet ablation and deposition depend sensitively on the background plasma conditions, in particular the electron temperature $T_e$, ion temperature $T_i$, and electron density $n_e$ (quantities that are often challenging to diagnose and reconstruct accurately), a complete predictive assessment of pellet fueling for full plasma shots should not rely on the pellet source model alone. The pellet source and the turbulent transport response must instead be treated self-consistently, since the evolving kinetic profiles affect the ablation process, while the pellet-driven perturbation in turn modifies the subsequent transport and profile evolution. This motivates embedding HPI2 within an integrated modeling framework. Similar coupled pellet-transport validation strategies have been pursued in previous studies, for example in JET pellet-fueling simulations \cite{Marin_2021}. These studies also showed that synthetic-diagnostic comparisons with interferometry provide a robust validation route for the coupled source-transport system. More broadly, using the TGLF turbulent transport model \cite{Staebler_2021_1,Staebler_2021_2}, successful full-radius L-mode integrated modeling studies have already demonstrated that this turbulent transport model can reproduce profile evolution in present devices \cite{Fonghetti_2025,angioni2022confinement}, providing an important basis for the present HPI2-TGLF modeling strategy.

Within the broader landscape of pellet modeling tools, the modeling workflow presented in this study is complementary to nonlinear MHD codes such as JOREK \cite{JOREK2021}, which excel at resolving local, fast dynamics including ablation, plasmoid expansion, and MHD response, but are not designed for long-timescale profile evolution under turbulent transport. Instead, this workflow targets the coupled source-transport problem on transport timescales, bridging the gap between such high-fidelity local simulations and lower-cost pulse-design or controller-oriented tools.

This article is organized as follows. Section~\ref{sec:hpi2_up} presents the HPI2 upgrade related to the spatial step and benchmarks against the previous implementation. Section~\ref{sec:val} reports a stand-alone validation of HPI2 against a pellet fueled WEST discharge. Section~\ref{sec:val_hfps} then demonstrates a fully predictive integrated modeling simulation of the same discharge, coupling the upgraded HPI2 to turbulent transport of heat and particles including impurities. Finally, section~\ref{sec:conc} summarizes the main conclusions and outlines future work.

\section{HPI2 Model Upgrade}\label{sec:hpi2_up}

As mentioned earlier, the pellet particle-source model considered here for stand-alone and integrated pellet modeling is HPI2 (Hydrogen Pellet Injection) \cite{Pegourie_2007_HPI2,koechl2012modelling}. HPI2 provides a computationally efficient description of the full chain from pellet entry into the confined plasma to the completion of plasmoid homogenization, yielding the particle deposition profile required by transport solvers. The model couples (i) an ablation module based on Neutral Gas and Plasma Shielding (NGPS) \cite{Pegourie_1993,Pegourie_2005} (with simplified neutral-gas shielding model NGS \cite{NGSmodel} used in low-temperature conditions to reduce computational cost) and including the influence of fast ions and electrons, and (ii) a homogenization module that evolves plasmoid expansion and the $\nabla B$-induced drift, and incorporates drift braking through Alfv\'en-wave-mediated magnetic tension as well as additional current-closure mechanisms (external and internal connections) that further reduce the drift depending on radial locations \cite{Sakamoto_2013}. In its standard implementation, HPI2 solves a set of 29 coupled differential equations to determine the redistribution of ablated material and the resulting source term. HPI2 has extensively been used in many devices, and can simulate any feasible pellet size and velocity.

In HPI2 the pellet path is discretized by advancing the pellet position in steps of size $dx$. At each step the model evaluates local background quantities from the provided equilibrium and profiles, computes the ablation rate and released material, and updates plasmoid dynamics and homogenization, iterating until complete ablation. The choice of $dx$ therefore sets the numerical resolution of the coupled ablation-homogenization dynamics and the representation of local gradients (including the proximity to rational surfaces), while also controlling computational cost and potentially numerical robustness. The key inputs are the magnetic equilibrium, safety factor $q$ profile, background electron temperature $T_e$, ion temperature $T_i$ and electron density $n_e$ profiles (and, where relevant, auxiliary-heating/fast-particle information), and pellet parameters (initial mass, composition, injection geometry and velocity). The main outputs are the particle deposition profile and, subsequently, the associated post-pellet electron density (and, when modeled, temperature) perturbation.

\begin{figure*}[!tbp]
    \centering
    \begin{subfigure}[t]{\linewidth}
        \centering
        \captionsetup{position=top} 
        \caption{Discretization parameter $dx$ scan.}
        \includegraphics[width=\linewidth]{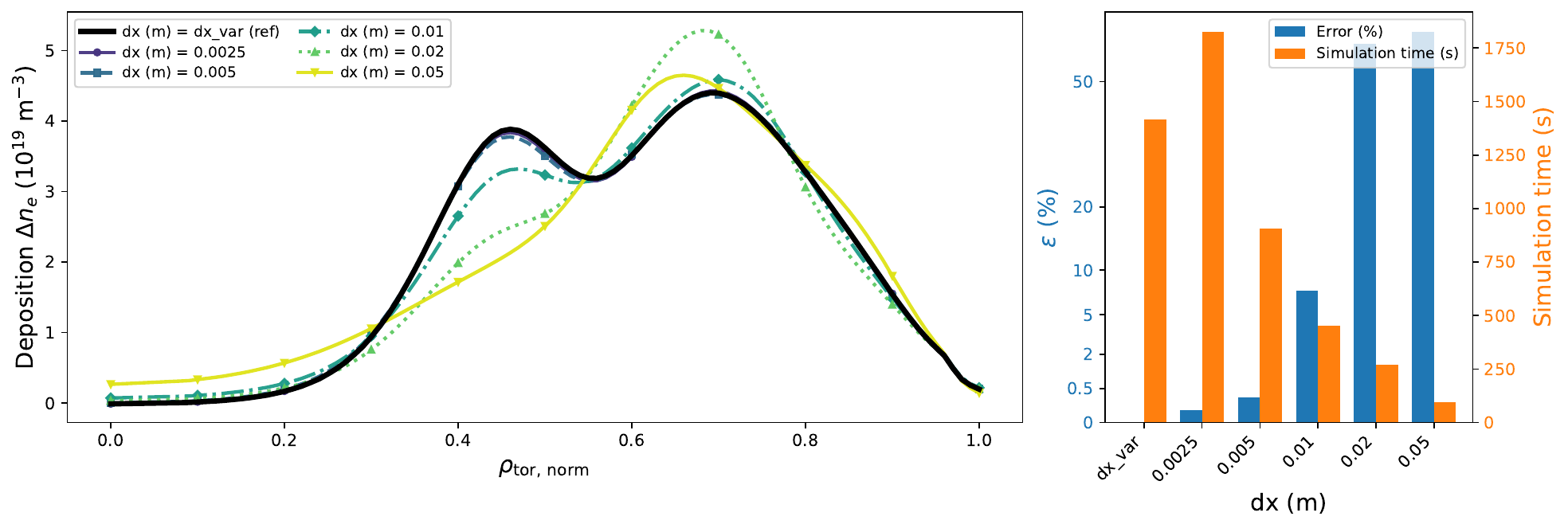}
        
        \label{fig:delta_ne_dx}
    \end{subfigure}

    \vspace{4pt}

    \begin{subfigure}[t]{\linewidth}
        \centering
        \captionsetup{position=top} 
        \caption{Scaling factor $A_{dx}$ scan.}
        \includegraphics[width=\linewidth]{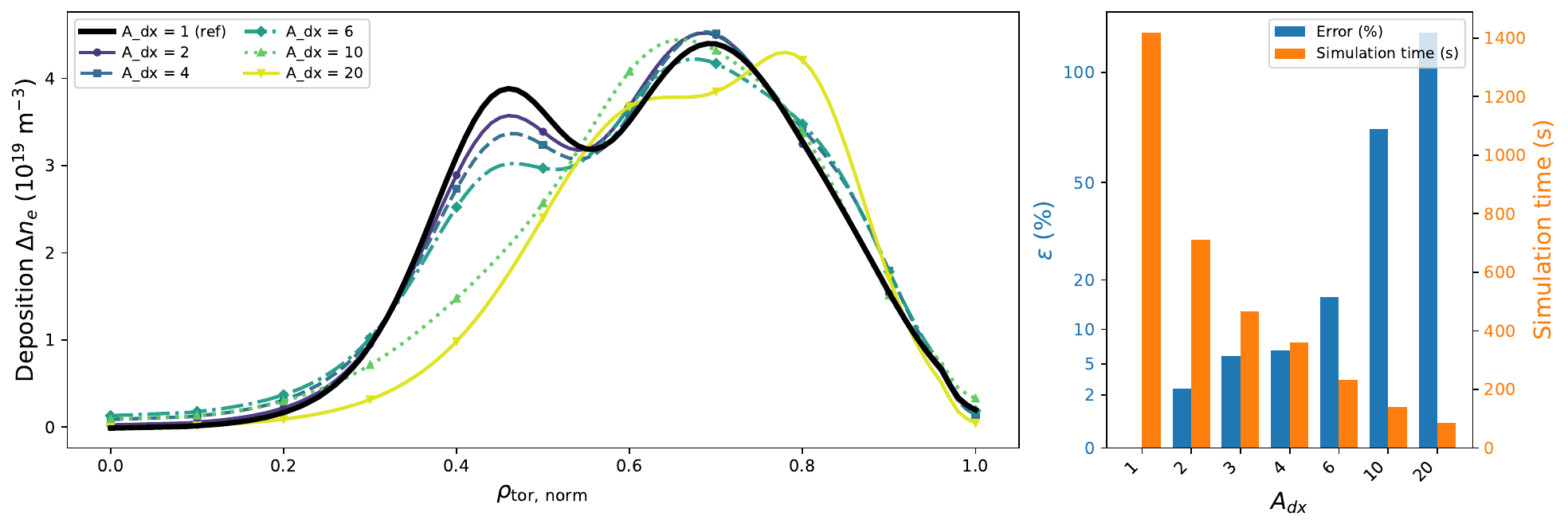}
        
        \label{fig:delta_ne_A}
    \end{subfigure}

    \caption{HPI2 density deposition profiles in a Low Field Side (LFS) injection case for WEST shot \#55589 (C4 campaign). \textbf{(a)}: sensitivity to the plasmoid release step, comparing several constant-$dx$ calculations with the self-consistent variable-step prescription $dx_{\mathrm{var}}$. \textbf{(b)}: sensitivity to the scaling factor $A_{dx}$ multiplying $dx_{\mathrm{var}}$, showing how increasing $A_{dx}$ progressively widens the discretization. In both panels, the reported error metric $\epsilon$ measures the mismatch with respect to the reference profile (HPI2 prediction using $dx=dx_{\mathrm{var,\: A_{dx}=1}}$)}, with emphasis on the deposition peak. The reported runtimes correspond to single-CPU executions.
    \label{fig:dx_A_subfig}
\end{figure*}

Although HPI2 contains a detailed description of ablation and homogenization, its numerical implementation depends on the spatial step $dx$ used to release successive plasmoids. In the standard version, $dx$ is prescribed by the user as a constant. A sensitivity scan shows that the predicted deposition profile can vary appreciably with this choice (figure~\ref{fig:delta_ne_dx}), indicating that $dx$ should not remain an arbitrary numerical parameter.

\begin{figure}[!htbp]
    \centering
    \includegraphics[width=\linewidth]{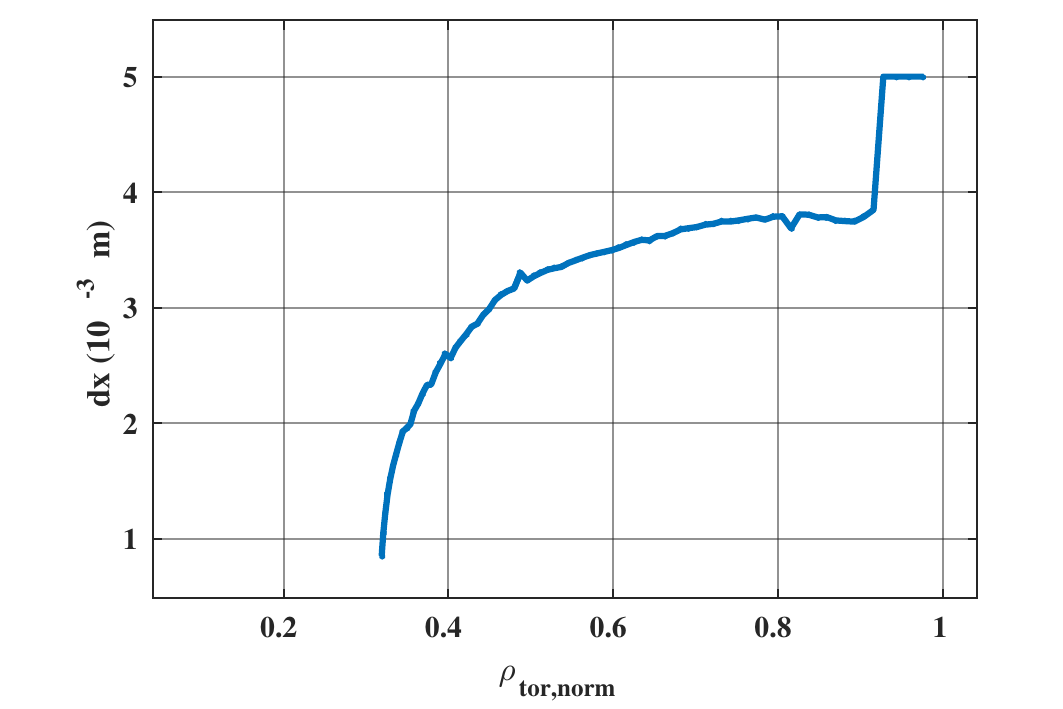} % <-- cambia el nombre/ruta
    \caption{HPI2 discretization parameter $dx$ with the new implementation, in a Low Field Side (LFS) injection case for WEST shot \#55589 (C4 campaign). At the edge the value of $dx$ is constant until the minimum $T_e$ is reached to calculate the plasmoids.}
    \label{fig:dx_rho}
\end{figure}

We therefore introduce a physics-based discretization by linking the step to the characteristic distance between successive plasmoid ejections. In HPI2 this distance is naturally associated with the time required for the pellet to emerge from the previously ejected plasmoid, $t_{\mathrm{exit}}$, which is computed self-consistently in the homogenization module (figure~\ref{fig:t_exit}). In practice, $t_{\mathrm{exit}}$ is the first time at which $|d_{\mathrm{plasmoid}}(t) - v_{\mathrm{drift}}\cdot t| \geq R_{\mathrm{plasmoid}}(t)$, where $d_{\mathrm{plasmoid}}(t)$ is the plasmoid centroid displacement, $v_{\mathrm{drift}}\cdot t$ is the pellet displacement in the same frame, and their difference gives the pellet-plasmoid separation. The plasmoid boundary is defined by the instantaneous cylindrical radius $R_{\mathrm{plasmoid}}(t)$, which is a dynamical variable of the homogenization ODE system that evolves under the combined effects of internal pressure and magnetic forces. $t_{\mathrm{exit}}$ therefore represents the time at which the plasmoid is considered to become fully detached from the pellet and a new ejection cycle can begin. The step is then defined locally as
  \begin{equation}
dx_{\mathrm{var}} \;=\; v_{\mathrm{pellet}}\cdot t_{\mathrm{exit}},
\end{equation}
where $v_{\mathrm{pellet}}$ represents the pellet velocity, assumed constant throughout the ablation process (unless the pellet rocket acceleration effect, assumed to be negligible in the cases described in this paper, is taken into account). Then $dx_{\mathrm{var}}$ becomes a variable quantity along the trajectory. This parametrization ties the numerical resolution to the underlying cyclic ablation-ionization-ejection picture and reduces sensitivity to an ad hoc choice of $dx$, while retaining the ability to resolve position-dependent effects such as interactions with nearby rational and semi-rational surfaces of the $q$ profile. In addition, using a self-consistent $dx$ allows the actual number of generated plasmoids to be determined, thereby providing a direct link to the corresponding deposition frequency. The resultant model for the plasmoid deposition frequency has been successfully validated at LHD \cite{pegourie2024structure}. In practice, the value of $dx_{\mathrm{var}}$ turns smaller when density and temperature raise in the background plasma, i.e. generally when the pellet gets closer to the core, then the accuracy of the model is higher (in the region where the pellet deposits more material). This can be observed in the Figure~\ref{fig:dx_rho}. Figure~\ref{fig:delta_ne_dx} compares the resulting deposition profiles with different values of constant-$dx$ with the new variable-step $dx_{\mathrm{var}}$, for the same WEST discharge. To quantify differences between pellet deposition profiles while emphasizing the physically relevant region of the deposition peak, we introduce a weighted error metric, denoted $\epsilon$, with respect to the reference deposition profile computed using the variable-step prescription $dx_{\mathrm{var}}$. For a given deposition profile $\Delta n_e(\rho)$ and the reference profile $\Delta n_e^{\mathrm{ref}}(\rho)$, the metric is defined as
 \begin{displaymath}
    \end{displaymath}
  \begin{equation}
    \epsilon =
    \frac{\sum_{\rho} w(\rho)\left[\Delta n_e(\rho)-\Delta n_e^{\mathrm{ref}}(\rho)\right]^2}{\sum_{\rho} w(\rho)\left[\Delta n_e^{\mathrm{ref}}(\rho)-\langle \Delta n_e^{\mathrm{ref}} \rangle_w\right]^2},
\end{equation}
where $\langle \Delta n_e^{\mathrm{ref}} \rangle_w = \sum_{\rho} w(\rho)\Delta n_e^{\mathrm{ref}}(\rho)$ denotes the weighted mean. In this form, $\epsilon = 0$ corresponds to perfect agreement, and larger values indicate larger mismatch with respect to the reference profile.
Here, $w(\rho)$ is a non-negative radial weighting function used to emphasize the physically relevant deposition region in the error metric. We define it as the normalized reference deposition profile,
\begin{equation}
w(\rho)=\frac{\Delta n_e^{\mathrm{ref}}(\rho)}{\sum_{\rho'}\Delta n_e^{\mathrm{ref}}(\rho')},
\end{equation}
so that $\sum_{\rho} w(\rho)=1$. With this definition, $w(\rho)$ has the same radial shape as $\Delta n_e^{\mathrm{ref}}(\rho)$ and only rescales it into relative weights. This choice makes the metric easy to interpret: the comparison is driven mainly by the deposition-peak region, which is the part most relevant for pellet-fueling predictions, while regions with negligible deposition do not dominate the error estimate.

Figure~\ref{fig:delta_ne_dx} shows that the predicted deposition profile can vary significantly when changing the discretization step $dx$, highlighting that $dx$ should not be treated as an arbitrary numerical parameter and motivating the physics-based variable-step formulation introduced here. Further details are provided in \cite{panera_alvarez_2026_19949047}.
To reduce computational cost, we additionally introduce a scaling factor $A$ such that $dx_{\mathrm{var}} \rightarrow A \cdot dx_{\mathrm{var}}$, with $A=1$ corresponding to the baseline variable-step model. Figure~\ref{fig:delta_ne_A} shows the impact of $A$ on the deposition profile. Values up to $A\simeq 4$ provide a useful speed-up with limited degradation relative to the reference case ($\epsilon<10\%$), whereas larger factors progressively modify the deposition and are only appropriate when prioritizing runtime over accuracy. Although these differences can be slightly lower or higher depending on the case simulated, for example with lower velocity we can accept higher values of $A$, also for higher values of mass. Very low pellet velocity can reduce $dx_{\mathrm{var}}$, while a high pellet mass increases the number of plasmoids that must be computed before complete ablation (for similar local ablation rates). Consequently, although $A_{dx}=1$ provides the highest accuracy, it can substantially increase the computation time in those cases. In ITER-like devices, despite the machine size, most of the ablation might happen close to the edge, so too large values of $A$ may not be advisable if density and temperature are not high enough to compensate with a small $t_{exit}$. Overall, $A\lesssim 4$ can be taken as a good rule of thumb for the trade-off between accuracy and runtime, without forgetting that $A=1$ is the best choice for good accuracy.

\section{HPI2 Stand-Alone Validation against a pellet-fueled WEST discharge}\label{sec:val}

This section validates the upgraded HPI2 model against a pellet-fueled WEST discharge, focusing on the density perturbation produced by each injection. The objective is to assess whether HPI2, including the new $dx$ parametrization, reproduces the experimentally measured variation in density after pellet injection in a WEST discharge.

\subsection{Experimental discharge and diagnostics}\label{sec:val_shot}
\begin{figure*}[!tbp]
    \centering
    \begin{subfigure}[t]{0.48\textwidth}
        \centering
        % Replace the placeholder below with the Te(0) vs time plot file.
        \captionsetup{position=top} 
        \caption{$T_e(0)$.}
        \includegraphics[width=\linewidth]{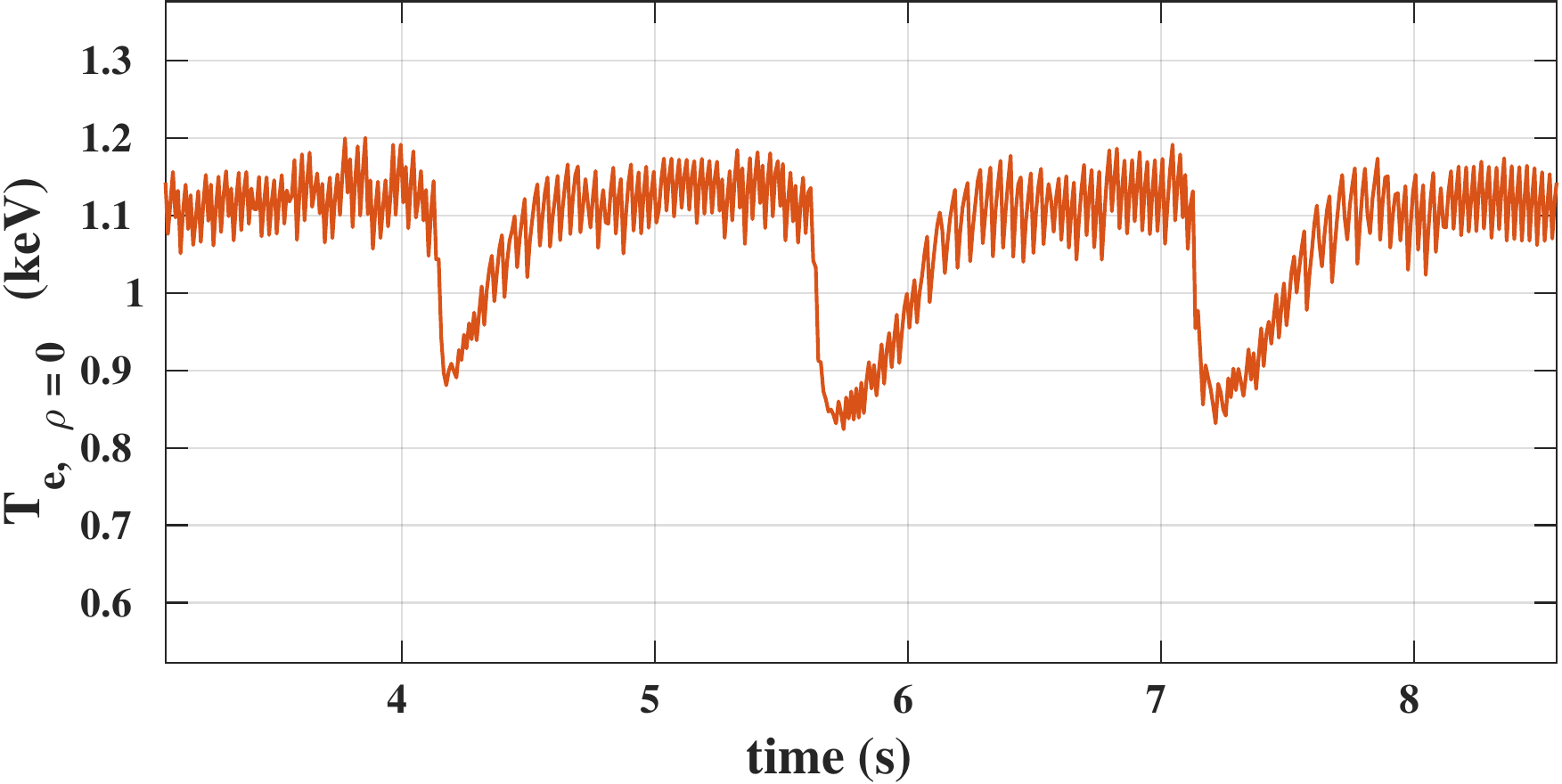}
        % \fbox{\parbox[c][0.22\textheight][c]{0.95\linewidth}{\centering Placeholder for $T_e(0)$ vs time}}
        
        \label{fig:val_overview_te0}
    \end{subfigure}\hfill
    \begin{subfigure}[t]{0.48\textwidth}
        \centering
        % Replace the placeholder below with the $P_{\mathrm{rad}}$ and $P_{\mathrm{ohmic}}$ vs time plot file.
        \captionsetup{position=top} 
        \caption{$P_{rad}$ and $P_{ohm}$.}
        \includegraphics[width=\linewidth]{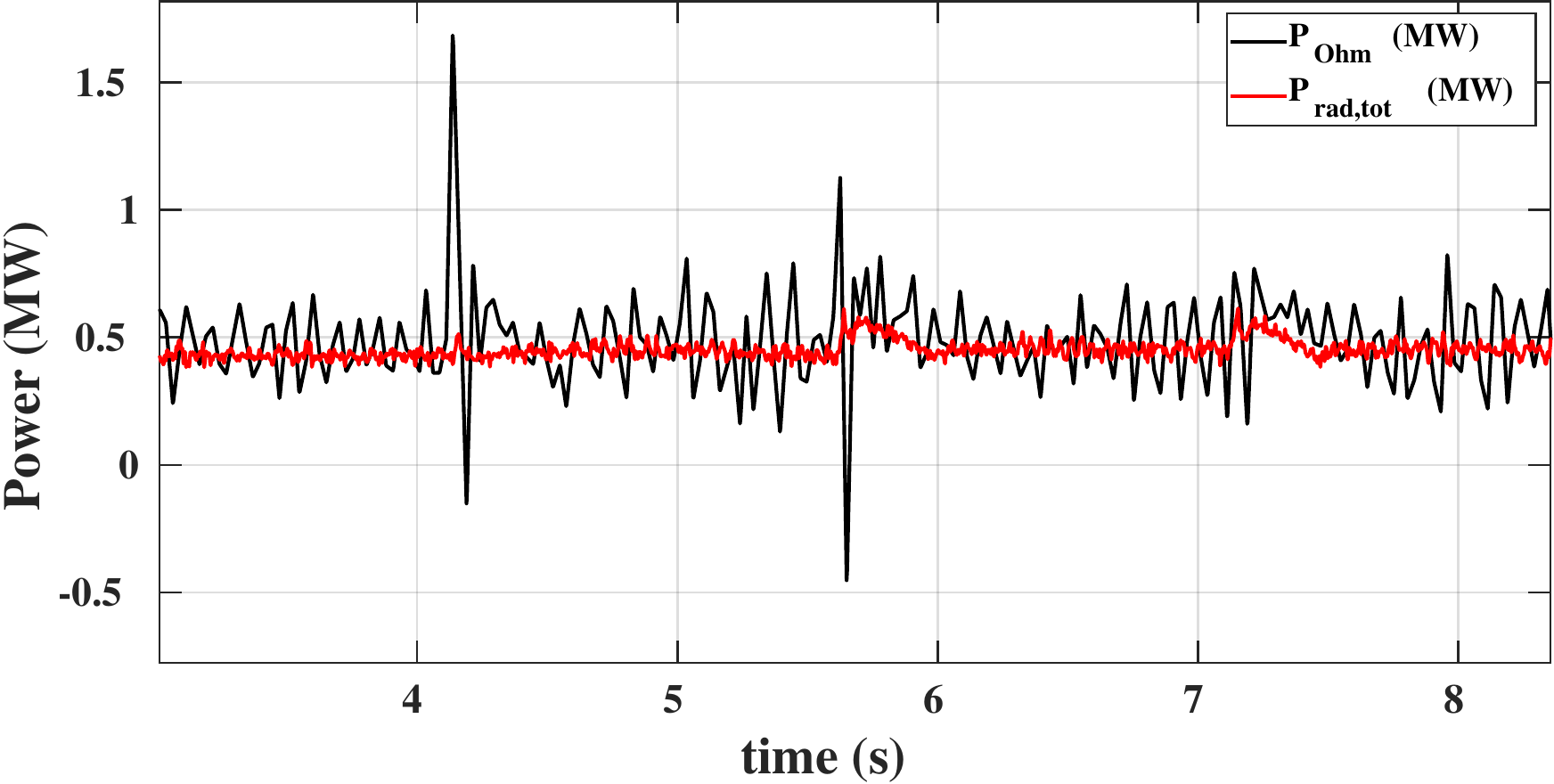}
        % \fbox{\parbox[c][0.22\textheight][c]{0.95\linewidth}{figures/sec 3/Pohm.pdf}}
        
        \label{fig:val_overview_power}
    \end{subfigure}

    \vspace{6pt}
    \begin{subfigure}[t]{0.35\textwidth}
        \centering
        \captionsetup{position=top} 
        \caption{WEST poloidal cut.}
        \includegraphics[width=\linewidth]{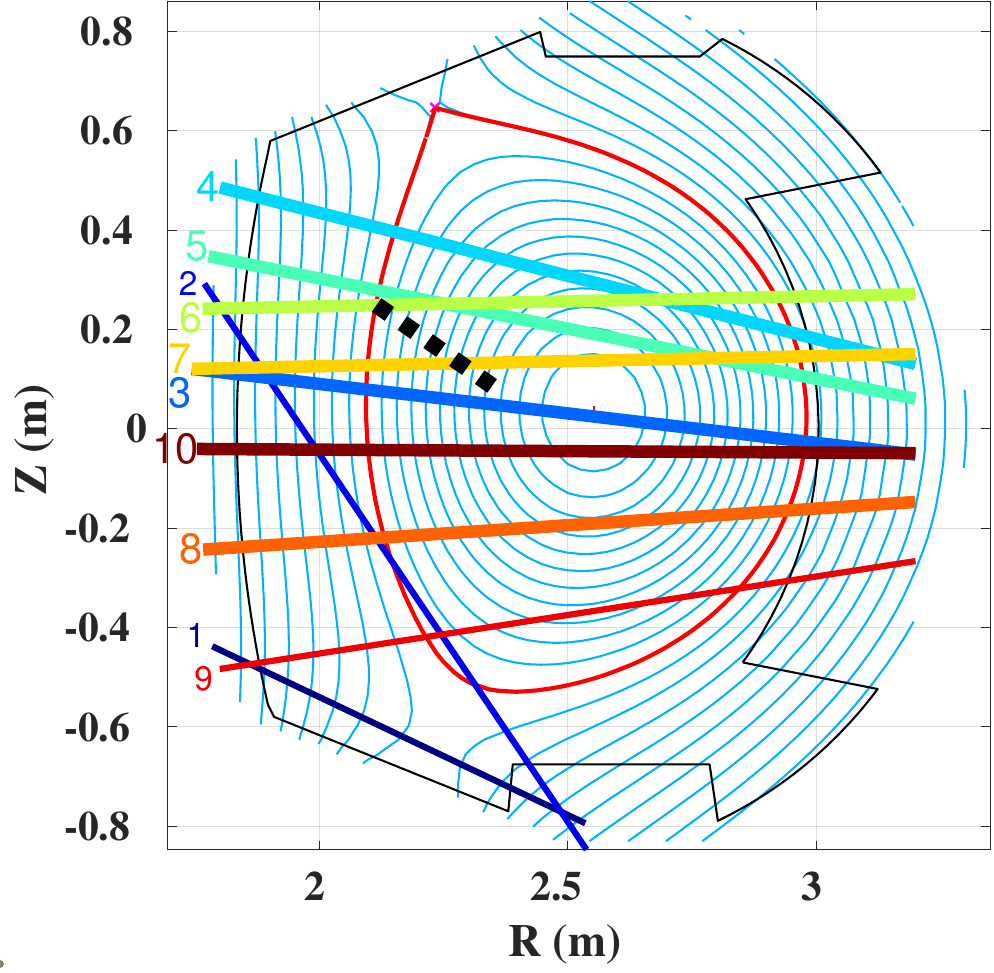}
        
        \label{fig:LoS}
    \end{subfigure}\hfill
    \begin{subfigure}[t]{0.60\textwidth}
        \centering
        \captionsetup{position=top} 
        \caption{$n_l$.}
        % Replace the placeholder below with the interferometry $n_l$ time traces.
        \includegraphics[width=\linewidth]{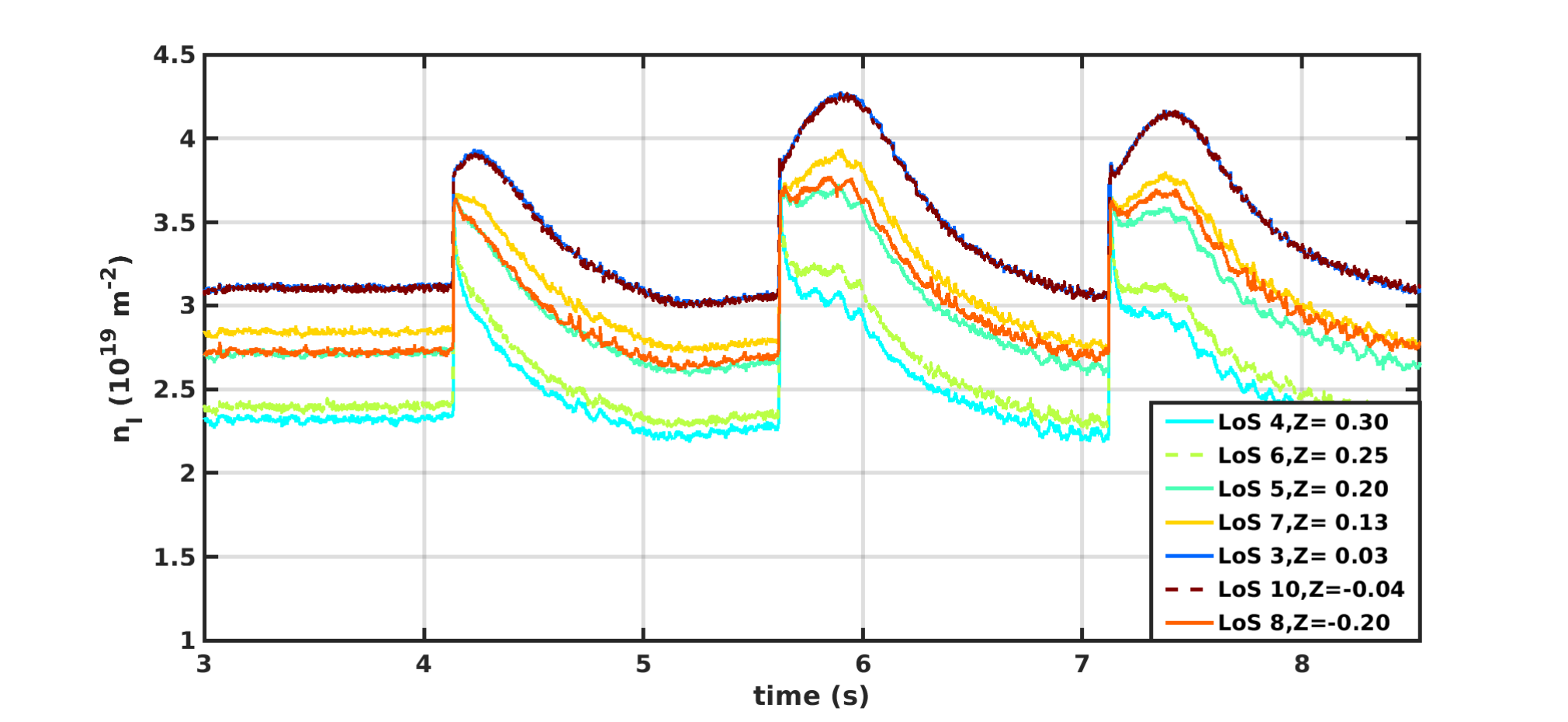}
        % \fbox{\parbox[c][0.22\textheight][c]{0.95\linewidth}{\centering Placeholder for interferometry line-integrated density $n_l$}}
        
        \label{fig:interferometry}
    \end{subfigure}

    \caption{Overview of the main diagnosed quantities at WEST discharge \#58656 during the flat-top: \textbf{(a)} electron temperature at the magnetic axis; \textbf{(b)} radiated and ohmic power; \textbf{(c)} WEST poloidal cut showing the interferometer lines of sight (numbered solid straight lines correspond to each LoS) and pellet trajectory (dashed black line), with wider lines corresponding to lines of sight used in this study; and \textbf{(d)} interferometry line-integrated density $n_l$ signals corresponding to the LoS used in the study.}
    \label{fig:val_overview}
\end{figure*}

The validation is performed on WEST discharge \#58656 (C7 campaign), an ohmic plasma (no auxiliary heating) with three deuterium pellets injected from the upper high-field-side (upHFS) injector \cite{alarcon2023versatile}. This L-mode ohmic shot was performed with a flattop toroidal current of around $I_p\approx0.5 $ MA, a toroidal magnetic field in the magnetic axis of $B_0=3.76$ T, and core electron temperature around $T_{e,\rho =0}=1.1\: keV$. Some other characteristic quantities of this discharge are shown in Figure~\ref{fig:val_overview}. The low pellet repetition rate (less than 1 Hz) allows the background plasma to recover between injections, enabling the three events to be analyzed independently. HFS injection is of particular interest because it generally provides deeper particle deposition and is therefore the preferred fueling geometry for ITER class devices \cite{Wiesen_2017}.

The main diagnostic used for validation is the WEST interferometer \cite{gil2019renewal}, which provides line-integrated density $n_l$ with sufficient temporal resolution ($\Delta t_{\mathrm{interferometry}} = 1\,\mathrm{ms}$) to capture the fast rise associated with pellet ablation and subsequent homogenization; line integrated density measurements from WEST interferometry are shown in Figure~\ref{fig:interferometry}, along with the respective line of sight (LoS) mapping in Figure~\ref{fig:LoS}. In this context, it is useful to distinguish timescales: the local ablation physics is very fast (typically $\mathcal{O}(0.1\,\mathrm{ms})$), while homogenization does not have a single characteristic time because it depends on transport conditions, combining fast parallel cloud expansion (also $\mathcal{O}(0.1\,\mathrm{ms})$) with slower cross-field redistribution driven by drifts and polarization. The effective buildup/redistribution of deposited matter is observed over a longer timescale, typically of order $\mathcal{O}(10\,\mathrm{ms})$. The reflectometer \cite{gil2019renewal} data are used to reconstruct the pre-pellet background density (together with interferometry); however, in this stand-alone validation they are not suitable for fully reconstructing the immediate post-pellet $n_e(\rho)$ profile because of their limited time resolution ($\Delta t_{\mathrm{reflectometry}} = 4\,\mathrm{ms}$) and the expected non-monotonic (often hollow) character of the post-pellet density, which challenges standard inversion methods. However, some qualitative comparisons are carried out to see the agreement of the HPI2 prediction with the measured reflectometry profile, shown in Figure~\ref{fig:refl_pellets}. The electron temperature $T_e$ is obtained from electron cyclotron emission diagnostic (ECE), $T_e$ at plasma core ($\rho_{tor,\, norm}=0$) is shown in Figure~\ref{fig:val_overview_te0}. The ion temperature $T_i$ is inferred from neutron detector. The magnetic equilibrium is reconstructed with NICE \cite{FAUGERAS2020112020} and is also used to compute synthetic interferometer signals from HPI2 outputs with Syndi \cite{Devynck2021}. The power is measured with bolometry, as shown in Figure~\ref{fig:val_overview_power}. It is worth pointing out that this is a fully ohmic discharge, with no external heating, and the radiated power in WEST is mainly coming from tungsten impurities.

Figure~\ref{fig:LoS} shows the ten interferometer lines of sight (LoS). We retain LoS 3-8 and 10, which sample the core/edge regions relevant to fueling, and exclude LoS 1 and 2 due to their strong edge/X-point sensitivity and reduced relevance for core fueling validation. LoS 9 is excluded as it shows a lack of consistency with the other interferometer lines of sight regarding the density profile reconstruction; this would introduce an additional reconstruction uncertainty that makes a quantitative comparison with HPI2 unreliable.

\subsection{Validation approach}\label{sec:val_method}
Direct validation against time-resolved local post-pellet $n_e(\rho,t)$ profiles is not feasible for the considered dataset, as explained in section \ref{sec:val_shot}. Instead, we validate HPI2 by comparing the measured line-integrated density  increment between pre- and post-pellet times,
\begin{equation}
\Delta n_l^{\mathrm{exp}} = n_l(t_{\mathrm{after}})-n_l(t_{\mathrm{before}}),
\end{equation}
with the corresponding synthetic diagnostic prediction computed from the HPI2 deposition result,
\begin{equation}
\Delta n_l^{\mathrm{HPI2}} = n_l^{\mathrm{syn}}(t_{\mathrm{after}})-n_l^{\mathrm{syn}}(t_{\mathrm{before}}).
\end{equation}
We define an accuracy metric for each LoS measurement
\begin{equation}
\alpha = \frac{\Delta n_l^{\mathrm{HPI2}}}{\Delta n_l^{\mathrm{exp}}},
\label{eq:alpha}
\end{equation}
so that $\alpha=1$ indicates perfect agreement. The times $t_{\mathrm{before}}$ and $t_{\mathrm{after}}$ are selected immediately before the rise and just at the post-pellet peak, respectively, to minimize sensitivity to slower transport effects not represented in HPI2.

The main uncertainties for modeling the pellet effect in this discharge are the pellet mass and velocity, the pre-injection background profiles, and the equilibrium reconstruction. First, the pellet mass and velocity delivered to the plasma can be substantially reduced by losses in the guide tube, an effect expected to be stronger for HFS injection due to the longer and more curved guide tube \cite{GERAUD20035,Geulin2023PhD_pelletInjection}. Second, uncertainties in the pre-injection profiles ($T_e$, $T_i$, $n_e$, $q$) and equilibrium reconstruction, affect the ablation rate, penetration and deposition drift. In this discharge, the core reflectometry and thomson scattering diagnostics were not available, which made the measured pre-injection and post-injection profiles be solely dependent in edge reflectometry and interferometry diagnostics.

% \subsection{}\label{sec:mass_vel}
Since direct pellet diagnostics were unavailable for the analyzed discharge, we constrain the pellet mass at plasma entry using the experimentally observed increase of total electron content immediately after injection, estimated via interferometry measurements. Assuming the rapid post-pellet rise is dominated by the deposited pellet particles, and the fact that fueling efficiency is close to 100\% in upper HFS pellet injection (due to the drift towards LFS), the inferred deposited particle number provides an effective estimate of the mass entering the separatrix. Applying this procedure to discharge \#58656, the inferred deposited particle number is converted into an injected mass (Table ~\ref{tab:pellet_params}) assuming one perturbed electron per deuterium atom and a solid deuterium density of $6.02\cdot 10^{28}$ atoms/m$^3$ at pellet-operating temperatures ($\sim 10\: K$). The resulting values indicate an approximate 50\% reduction relative to the nominal pellet mass launched by the injector(corresponding with $4.7\cdot10^{-9}m^3$ in volume). This mass estimate is used as input for HPI2 in the subsequent analysis. 

\begin{figure}[!htbp]
    \centering
    \includegraphics[width=\linewidth]{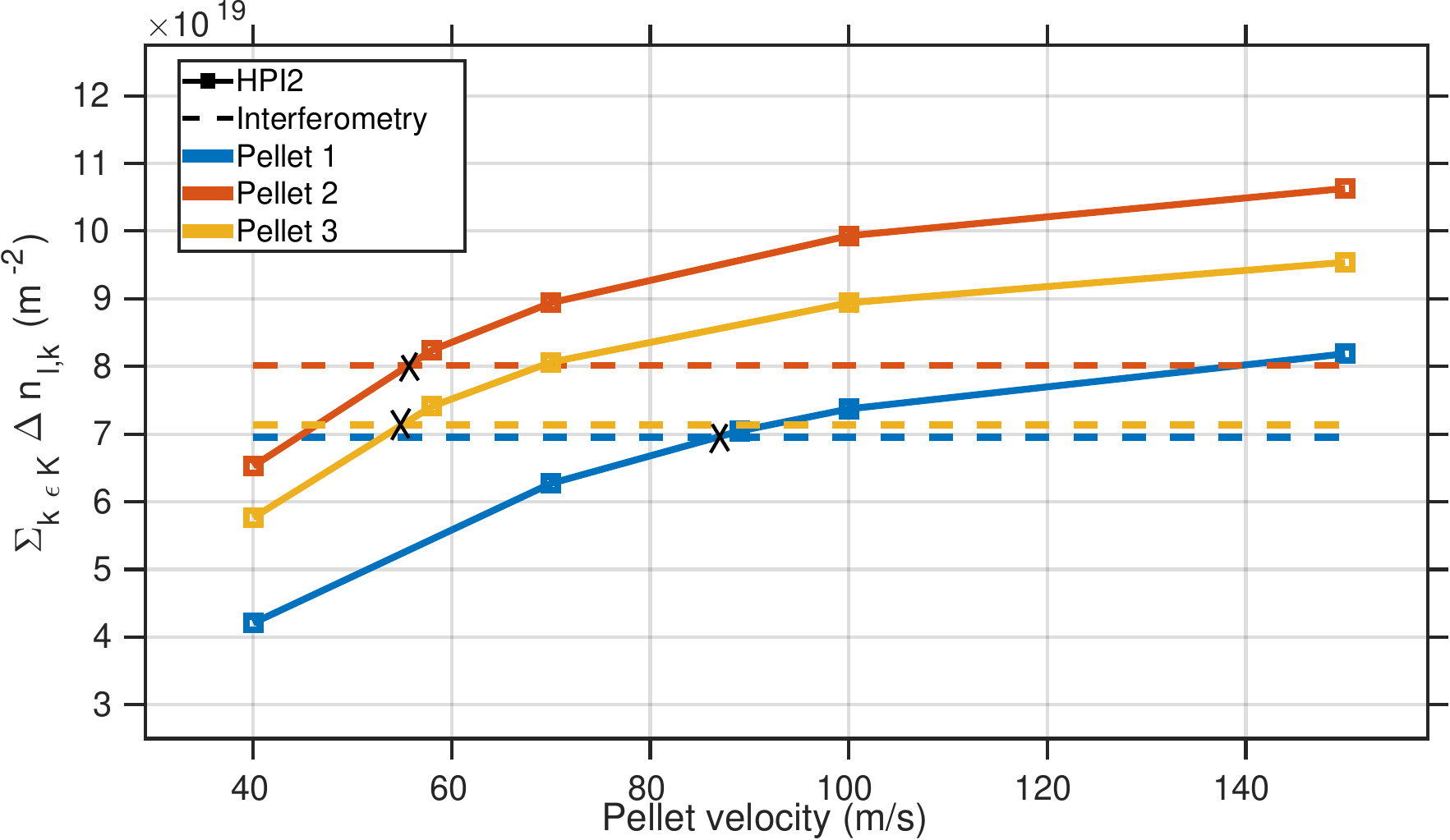} % <-- cambia el nombre/ruta
    \caption{Velocity scan used to constrain the pellet injection speed in WEST \#58656. The vertical axis represents the sum over the selected interferometry lines of sight (LoS) of the pellet-induced line-integrated density increment $\sum_{k \epsilon K} \Delta n_{l,k}$ with $K=\{3,4,5,6,7,8,10\}$ (in m$^{-2}$). The horizontal axis shows the pellet velocity $v_{\mathrm{pellet}}$. Squares represent HPI2 predictions $\sum_{k \epsilon K} \Delta n_{l,k}^{\mathrm{HPI2}}$ joined by solid lines. Dashed horizontal lines represent the interferometry measurement  $\sum_{k \epsilon K} \Delta n_{l,k}^{\mathrm{exp}}$ for each pellet. First pellet is shown in blue, second in red, and third in yellow. The X markers represent the intersection for the velocity estimate.}
    \label{fig:velocity_scan_sumLos}
\end{figure}

Because the pellet velocity cannot be determined accurately for this discharge (no fast camera and no dedicated pellet diagnostics were available), we perform a velocity scan around plausible values. Each HPI2 run uses the pellet mass inferred above and the remaining inputs taken from the reconstructed equilibrium/profiles and measurements immediately before the pellet enters the plasma. The plausible velocity range is inferred from the ablation duration observed in the interferometry signals for each pellet, assuming a penetration depth between 20~cm (lower bound) and 50~cm (WEST minor radius). The measured ablation durations are 3~ms, 5~ms and 5~ms for the first, second and third pellets, respectively, yielding a first order approximation of $v_\mathrm{pel}\simeq 66\,\mathrm{m/s}$-$166\,\mathrm{m/s}$ for the first pellet and $40\,\mathrm{m/s}$-$100\,\mathrm{m/s}$ for the second and third pellets. Accordingly, the scan is performed using HPI2 over $v_\mathrm{pel}=40\,\mathrm{m/s}$ to $150\,\mathrm{m/s}$, while the nominal injector setpoint velocity is $350\,\mathrm{m/s}$ for this discharge. The expected velocity arriving to the tokamak chamber in upper HFS injection is in agreement with the estimations of $120-200\,\mathrm{m/s}$ from \cite{GERAUD20035} in Tore Supra. The results of this scan are summarized in Figure~\ref{fig:velocity_scan_sumLos}. For each assumed pellet velocity, the HPI2 model predicts the sum of the pellet-induced line-integrated density increments over the selected interferometry lines of sight ($\sum_{k \epsilon K} \Delta n_{l,k}$ with $K=\{3,4,5,6,7,8,10\}$). This quantity decreases as the pellet velocity is reduced. The experimentally measured values for each pellet are shown as horizontal reference lines, and their intersection with the HPI2 curve defines the estimated pellet velocity reported in Table~\ref{tab:pellet_params}. The different lines of sight have equal weight in the sum, which implicitly gives higher weight to those lines of sight where the pellet induced $\Delta n_l$ is more pronounced.

It is important to emphasize that this decrease in the simulated signal does not imply a reduction in the total number of deposited electrons. The summed line-integrated density is a diagnostic-specific quantity, based on a limited set of interferometer chords, and therefore samples only part of the plasma volume rather than providing a full tomographic reconstruction. As a result, changes in the spatial distribution of the deposition (driven by the pellet velocity) can modify the measured signal without corresponding to a proportional change in the total particle inventory.

In Table~\ref{tab:pellet_params}, the difference between the three pellets is noticeable, in both mass and velocity losses. First pellet losses are higher in mass, while lower in velocity, than second and third pellets. This could be related to the fact that the first pellet is launched into a clean guide tube, while the second and third pellets encounter some residual material from the first pellet that can decrease their velocity and thereby reduce mass losses. The friction of the pellet on the wall of the guide tube, along with the fact that the WEST pellet injector is not in routine use, may contribute to the system not reaching its nominal design performance. This might help to explain the difference between the expected velocities by pellet injector design of $120-200\,\mathrm{m/s}$ mentioned in \cite{GERAUD20035}, and the obtained estimated velocities of $59-89\,\mathrm{m/s}$ in this study.

\begin{table}[!htbp]
  \centering
  \small
  \setlength{\tabcolsep}{6pt}
  \begin{tabular}{|p{5cm}|c|c|c|}
  \hline
  \textbf{Pellet} & $1^{\mathrm{st}}$ & $2^{\mathrm{nd}}$ & $3^{\mathrm{rd}}$ \\ \hline
  \textbf{Electron perturbation} $(10^{19})$ & 13.2 & 16.2 & 14.8 \\ \hline
  \textbf{Spherical radius} (mm) & 0.81 & 0.86 & 0.84 \\ \hline
  \textbf{Volume} $(10^{-9}\,\mathrm{m}^3)$ & 2.23 & 2.66 & 2.48 \\ \hline
  \textbf{Mass loss} (\%) & 53 & 43 & 47 \\ \hline
  \textbf{Velocity} ($\mathrm{m/s}$) & 89 & 58 & 58 \\ \hline
  \textbf{Velocity loss} (\%) & 75 & 83 & 83 \\ \hline
  \end{tabular}
  \caption{Estimated pellet mass and velocity for WEST discharge \#58656 using interferometry data and HPI2 predictions.}
  \label{tab:pellet_params}
\end{table}
\subsection{Results and discussion}\label{sec:val_results}

Using the estimated pellet mass and pellet velocity from Table~\ref{tab:pellet_params}, and the corresponding HPI2 inputs for each of the pellets in the studied discharge, we predicted the pellet deposition profile with HPI2 and compared them to the interferometry data using the figure of merit $\alpha$ defined in equation \ref{eq:alpha}. Overall, Table~\ref{tab:alpha} shows that $\alpha$ remains close to unity for all available interferometer chords and for the three consecutive injections, supporting a quantitatively robust match between the predicted and measured line-integrated density increments. The chord-to-chord spread is moderate (pellet~1: $\alpha\in[0.84,1.35]$; pellet~2: $\alpha\in[0.85,1.29]$; pellet~3: $\alpha\in[0.94,1.23]$), and the mean deviation decreases from $\langle|\alpha-1|\rangle=13.1\%$ and $12.3\%$ for pellets~1 and 2 to $6.6\%$ for pellet~3. The largest deviations are found in LoS~8, one of the outer chords (see Figure~\ref{fig:LoS}), where HPI2 clearly overpredicts. However, LoS 4 seems to be equally far from the magnetic axis and its accuracy is higher, so the discrepancies could be due to the equilibrium reconstruction uncertainty rather than due to an edge-related effect. In LoS~3 and 10, which pass almost directly through the magnetic axis (see Figure~\ref{fig:LoS}), slightly underprediction is observed. A bit farther from the magnetic axis, we find LoS~4, 5, and 6, where HPI2 slightly overpredicts. This suggests that HPI2 slightly overpredicts the edge response and underpredicts the core response. It is also worth noting that the inferred pellet velocity is not unique: different values within the plausible range can provide similarly good agreement. In addition, the interferometry signals used to constrain the pellet velocity carry their own experimental error bars, and the chord-by-chord comparison relies on NICE equilibrium reconstruction, which introduces additional uncertainty. Therefore, the estimated pellet velocity should be interpreted within the combined measurement and reconstruction uncertainties. In this context, the absence of a simple correlation between the estimated velocities/masses and the accuracy metrics is not unexpected (see Table~\ref{tab:pellet_params} and Table~\ref{tab:alpha}). However, this level of agreement across multiple, independent lines of sight indicates that HPI2 captures the correct local magnitude of the pellet particle source (within the uncertainties of the inferred pellet mass/velocity, pre-injection profiles and equilibrium reconstruction), and therefore provides a solid stand-alone validation baseline before moving to the integrated modeling validation in Section~\ref{sec:val_hfps}.

The agreement achieved for the three independent injections provides a first experimental validation of HPI2 in the WEST tokamak, supporting the applicability of the code with the upgraded $dx$ treatment. It is worth mentioning that comparisons with the previous fix $dx$ scheme are not shown since HPI2 predictions with low enough fix $dx$ values (at higher computational cost) tend to overlap $dx=dx_{var}$ predictions, as it can be appreciated in Figure~\ref{fig:delta_ne_dx}. Additionally, the main objective of the upgraded $dx$ technique is improving the robustness and physical meaning of the discretization \cite{pegourie2024structure}.
\begin{table}[!htbp]
\centering
\begin{tabular}{|l|l|l|l|}
\hline
\textbf{Pellet}  & 1st  & 2nd  & 3rd   \\ \hline
\textbf{LoS 3}   & 0.84 & 0.85 & 0.94  \\ \hline
\textbf{LoS 4}   & 1.03 & 1.05 & 1.01  \\ \hline
\textbf{LoS 5}   & 1.02 & 0.99 & 0.96  \\ \hline
\textbf{LoS 6}   & 1.08 & 1.17 & 0.94  \\ \hline
\textbf{LoS 7}   & 0.84 & 0.94 & 1.004 \\ \hline
\textbf{LoS 8}   & 1.35 & 1.29 & 1.23  \\ \hline
\textbf{LoS 10}  & 0.88 & 0.87 & 0.94  \\ \hline
\textbf{Avg. $|\alpha-1|$ (\%)} & 13.1 & 12.3 & 6.6 \\ \hline
\textbf{Minimum} & 0.84 & 0.85 & 0.94  \\ \hline
\textbf{Maximum} & 1.35 & 1.29 & 1.23  \\ \hline
\end{tabular}
\caption{Figure of merit $\alpha$ measured for every line of sight (LoS) of the WEST interferometer for the discharge \#58656. Accuracy of the HPI2 model using the estimated pellet mass and velocity.}
\label{tab:alpha}
\end{table}
\begin{figure*}[!tbp]
\centering
\begin{subfigure}[t]{0.33\textwidth}
  \centering
  \includegraphics[width=\linewidth]{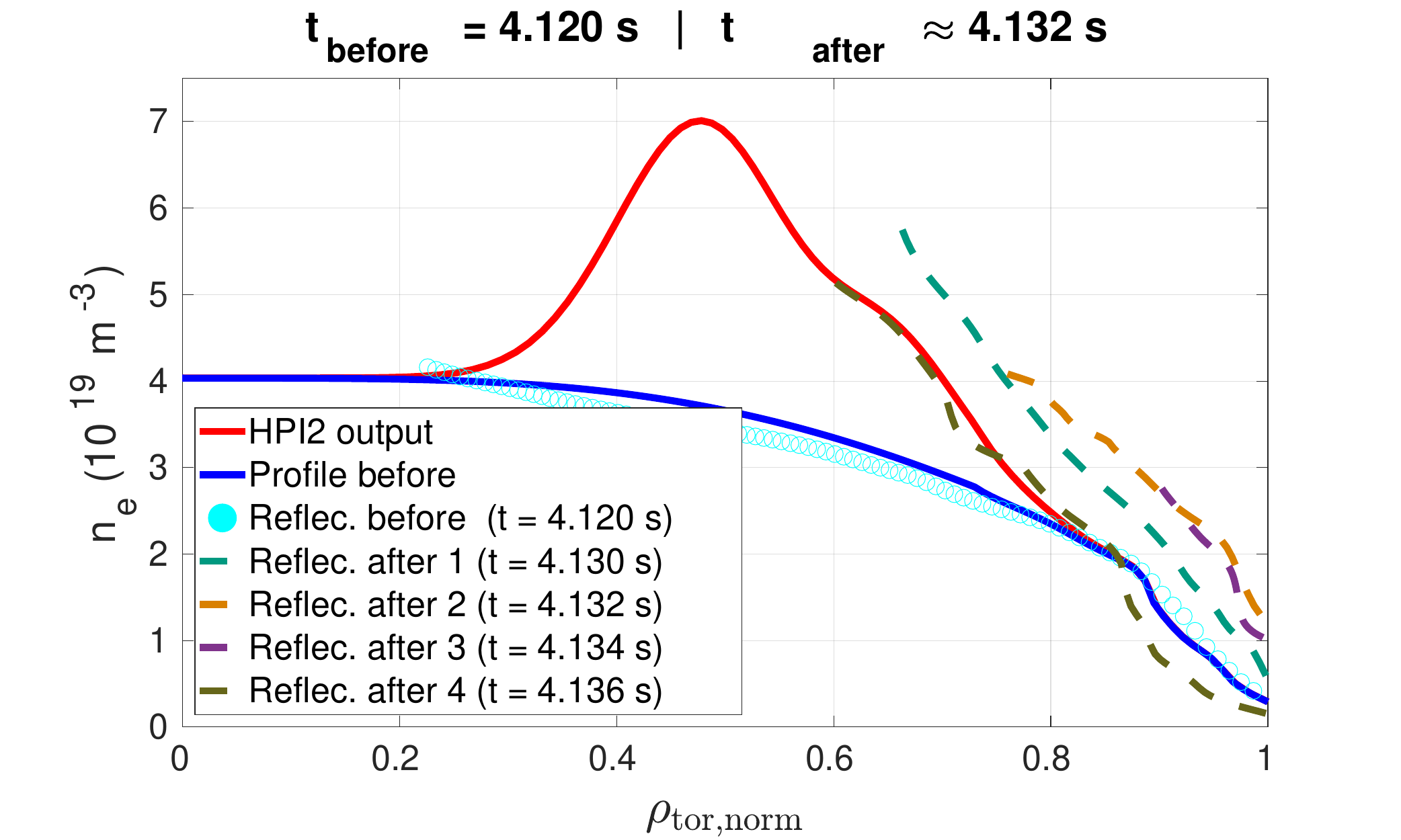}
  \caption{Pellet 1}
  \label{fig:refl_pel1}
\end{subfigure}\hfill
\begin{subfigure}[t]{0.33\textwidth}
  \centering
  \includegraphics[width=\linewidth]{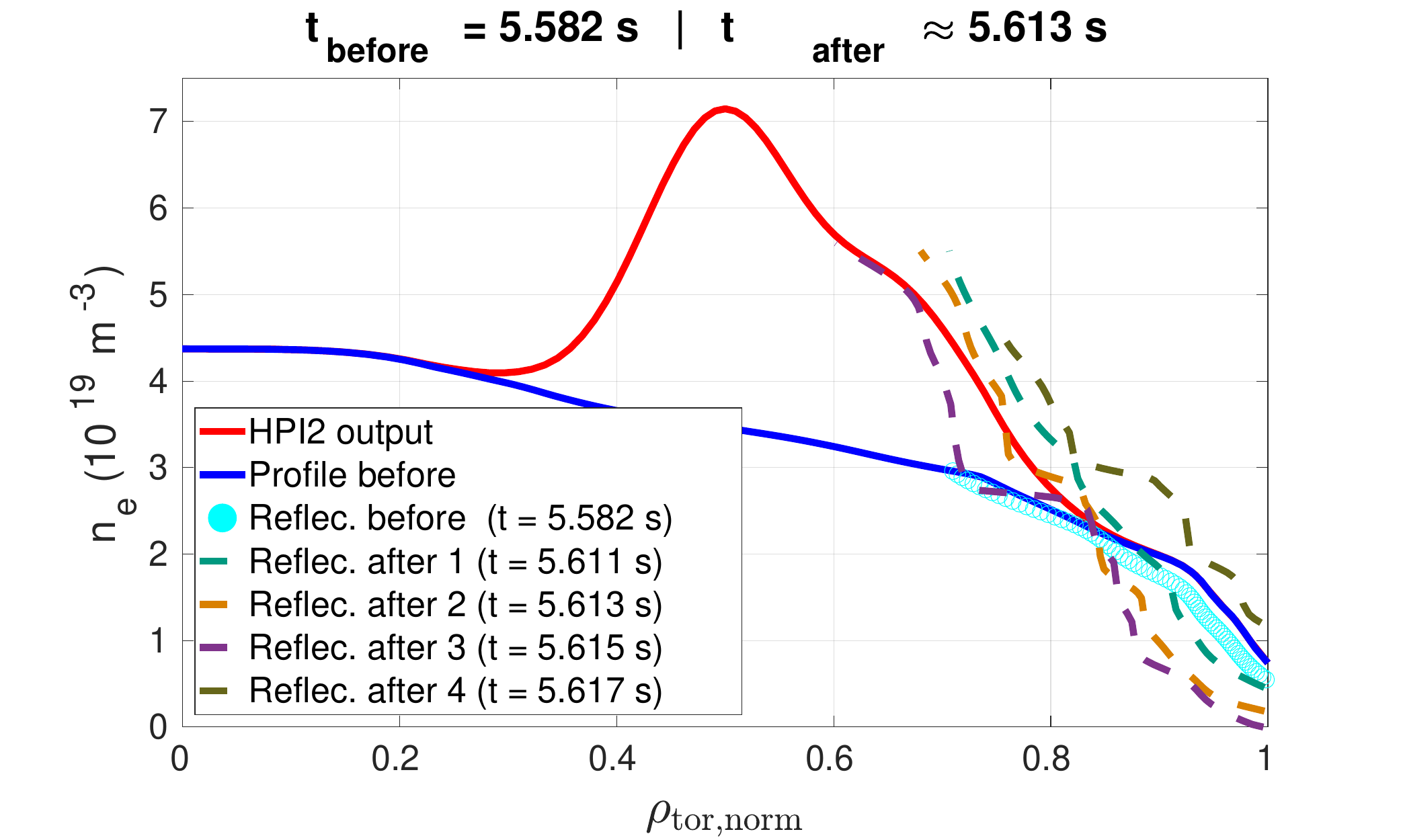}
  \caption{Pellet 2}
  \label{fig:refl_pel2}
\end{subfigure}\hfill
\begin{subfigure}[t]{0.33\textwidth}
  \centering
  \includegraphics[width=\linewidth]{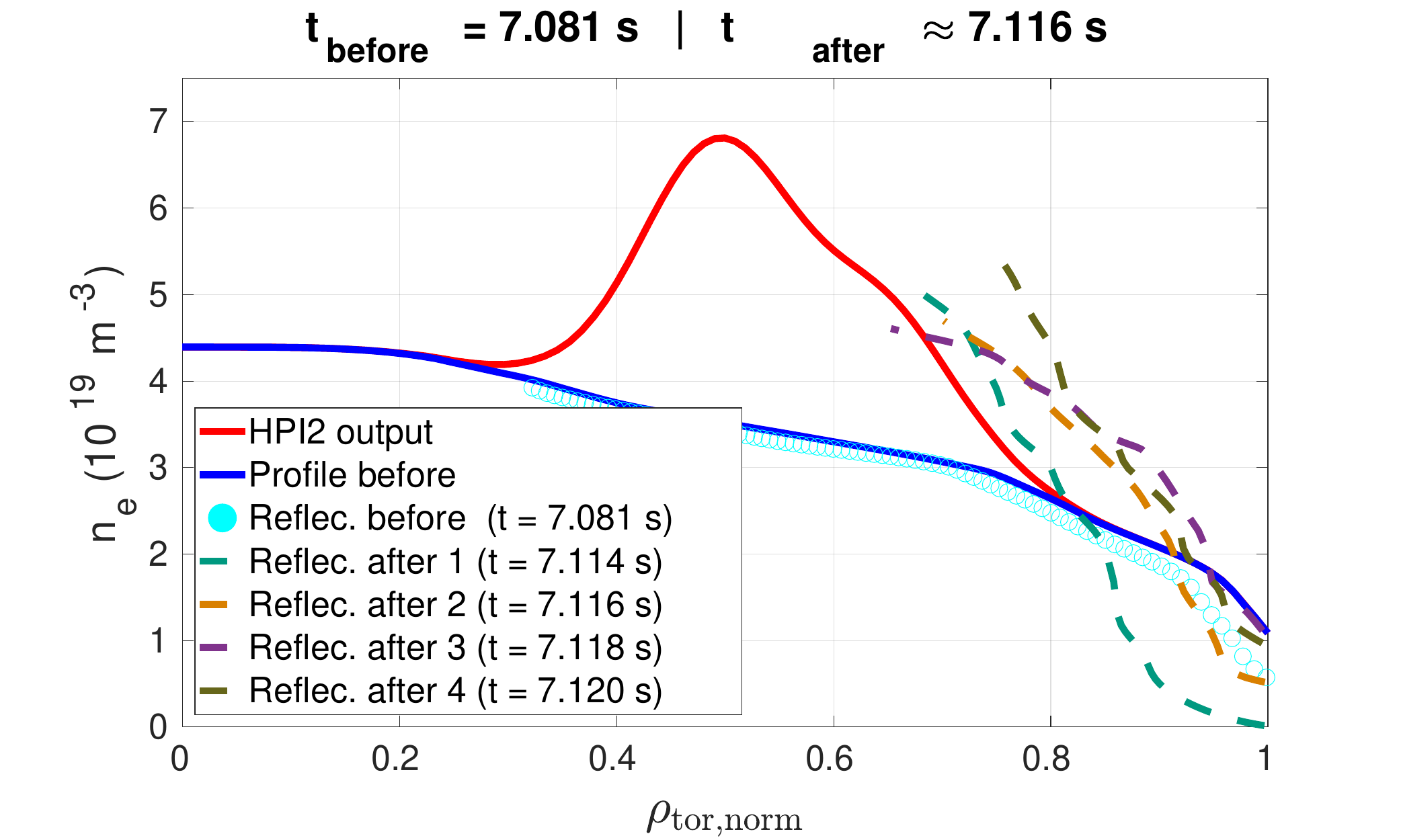}
  \caption{Pellet 3}
  \label{fig:refl_pel3}
\end{subfigure}
\caption{Reflectometry measured density profiles immediately before (blue circles) and just after (dashed lines) each pellet injection, compared with the $n_e(\rho)$ profile (reconstruction using interferometry and reflectometry data) used as input to HPI2 and the post-pellet $n_e(\rho)$ output predicted by HPI2.}
\label{fig:refl_pellets}
\end{figure*}

To back this validation, Figure~\ref{fig:refl_pellets} shows, for each of the three injections, the reflectometry density profile before (according to the time used for the HPI2 simulation), and just after the pellet (according to available reflectometry data), compared with the $n_e(\rho)$ profile (combining reflectometry and interferometry data) used as input to HPI2 and the resulting post-pellet profile predicted by HPI2. The agreement is overall good, and in particular the density gradient is consistently reproduced in the last data points of the reflectometer towards the core.

However, reflectometry cannot be trusted on its own for a quantitative validation because radial shifts can occur (depending on the inversion and on the fast transient conditions after the pellet), and it does not provide the profile up to the magnetic axis. To illustrate this radial shifts it can be seen how the edge reflectometry profiles show on Figure~\ref{fig:refl_pellets} lower density after the pellet in the edge region in some time steps, which is counterintuitive since the pellet represents a particle source. However, due to the radial shift, in other time steps this is not the case. For this reason, we interpret this comparison as a qualitative consistency check, and we put attention towards correct gradients and overall shape rather than a perfect fit. Together with the quantitative validation obtained from interferometry (Table~\ref{tab:alpha}), it provides a complementary confirmation that the upgraded HPI2 model is validated on WEST \#58656.

It is worth mentioning that in the simulated cases, the plasmoid drift is small in absolute terms, typically $\lesssim 5\, cm$, compared to WEST minor radius ($a=50\, cm$). The main reason may be the low $T_e$ (in this ohmic pulse) on which the drift displacement depends (this is because lower Te gives a lower ablation rate and a smaller initial plasmoid pressure to which the drift acceleration is proportional to first order). The magnetic field might also cause a stronger reduction of the drift if it is relatively high (as it happens when injecting from the high field side) as the drift damping terms increase with $B$. More detailed analysis on the plasmoid drift dependencies can be found in \cite{Pegourie_2007_HPI2,Panadero_2023}. Ablation in the very periphery $\rho_{tor,norm}\gtrapprox 0.85$ is neglected in the calculations shown in Figure~\ref{fig:refl_pellets} as the contribution of ablation at very low background temperatures in that region (compared to interior regions at higher temperature) is expected to be very small due to the strong dependence of the ablation on $T_e$.

\section{Validating pellet source and transport self-consistently using high fidelity integrated modeling.}\label{sec:val_hfps}

The previous section provides a constrained estimate of the pellet parameters (mass and injection velocity) and a first assessment of the associated experimental uncertainties. In this section, we move to fully time-dependent, full-radius integrated modeling to validate HPI2 when coupled to the High Fidelity Pulse Simulator (HFPS), i.e. a tokamak-agnostic workflow originating from JINTRAC \cite{romanelli2014jintrac} and relying on the IMAS I/O standard. In this configuration, the interplay between pellet fueling, multi-channel turbulent transport and impurity/radiation dynamics is treated self-consistently over the full discharge evolution, thus providing a validation of the interplay between turbulent transport and pellet source.

\subsection{Integrated modeling set-up and predictive scope}\label{subsec:hfps_setup}
The integrated modeling exercise has been carried out using HFPS as the 1.5D core transport solver. Turbulent heat and particle transport is modeled with TGLF employing the SAT2 saturation rule \cite{Staebler_2021_1}\cite{Staebler_2021_2}, while neoclassical transport is modeled with NCLASS \cite{houlberg1997bootstrap}. The neutral particle source is provided by FRANTIC \cite{tamor1981antic}, and pellet ablation and deposition are modeled with HPI2 \cite{Pegourie_2007}\cite{koechl2012modelling} (including the new improvement), using the pellet mass and injection velocity constrained by the stand-alone validation against WEST \#58656 interferometry. Impurity transport and radiation are modeled with SANCO \cite{romanelli2014jintrac}, introducing nitrogen (N) and tungsten (W) as the dominant impurities in this W-wall device.

To better understand the coupling of the HPI2 model within the HFPS framework, it is worth mentioning that HPI2 is called once per pellet injection event, using a single time slice of plasma quantities and profiles to calculate the deposition profile, and the combined ablation and homogenization time that is used in HFPS. This means that the HPI2 output does not take into account the evolution of the plasma profiles due to transport during the few milliseconds that this process takes. To implement this in HFPS, the modeling choice has been to divide the deposition profile by the ablation and homogenization time, and introduce this profile as a source during the corresponding time range. This is a key distinction from a simpler workflow in which HPI2 would be run externally, and its output imported as a static density increment applied instantaneously: by distributing the source in time, the coupled transport, impurity, and radiation modules continue to evolve concurrently with particle deposition, so that the plasma response is captured during the pellet ablation and homogenization event rather than only after it.

As previously mentioned, the WEST discharge \#58656 is only heated ohmically, and therefore does not include external auxiliary heating (NBI/ICRH/ECRH/LH) which reduces model uncertainties and facilitates the validation of the pellet response. Sawtooth activity is described with the Kadomtsev reconnection model applying a periodic sawtooth reconnection frequency. At the plasma edge, the separatrix boundary conditions are imposed and kept fixed in time, with $T_{e,\mathrm{sep}} = T_{i,\mathrm{sep}} = 20\,\mathrm{eV}$, equal to the neutrals temperature (via gas injection), and $n_{e,\mathrm{sep}} = 0.5\times10^{19}\,\mathrm{m}^{-3}$, which provides a sufficiently representative boundary conditions for the present ohmic discharge, while also constituting a limitation for the quantitative description of the edge plasma. In addition, the concentrations of the selected impurities, nitrogen and tungsten, are adjusted through feedback control on the separatrix N and W density: the nitrogen boundary influx is tuned to reproduce the experimental $Z_{\mathrm{eff}}$, while the tungsten boundary influx is tuned to match the measured radiated power $P_{\mathrm{rad}}$. With these choices, the validation focuses on the coupled response of (i) the HPI2 pellet particle source, (ii) turbulent transport (TGLF-SAT2), and (iii) impurity/radiation dynamics (SANCO) in reproducing the experimentally observed density and temperature evolution following pellet injection.

This choice is motivated by the demonstrated capability of physics-based integrated modeling to reproduce full-radius kinetic profile dynamics, and by the recognized role of impurity radiation (especially in W environments) in shaping the ohmic temperature evolution \cite{bourdelle2025integrated, angioni2022confinement}.

The simulation window covers the full flat top discharge phase of interest, including the pre-pellet baseline and the post-pellet relaxation for the three pellets that are injected, and is initialized from experimental equilibrium reconstruction and reconstructed kinetic profiles at time $t=3\, s$ (chosen sufficiently before the pellet injection to minimize sensitivity to initial conditions). The pellet is injected from the upper high-field side injector, and the pellet parameters of mass and velocity are fixed to the values identified as most suitable in the stand-alone validation in table~\ref{tab:pellet_params}.

A compact summary of the modeling ``duties'' (predictive vs prescribed quantities and associated modules) is given in Table~\ref{tab:hfps_duties}, following good practices in integrated modeling validation studies.

\begin{table*}[!tbp]
\centering
\caption{HFPS ``duties table'' for WEST \#58656. ``Predictive'' denotes self-consistent time evolution in the simulation; ``Prescribed'' denotes imposed from experiment/reconstruction.}
\label{tab:hfps_duties}
\small
\setlength{\tabcolsep}{4.5pt}
\renewcommand{\arraystretch}{1.12}

\begin{tabular*}{\textwidth}{@{\extracolsep{\fill}} l p{2.35cm} p{2.55cm} p{2.55cm} p{3.05cm} p{3.05cm} @{}}
\toprule
\textbf{Duty} & \textbf{Framework} & \textbf{Current profile} & \textbf{Equilibrium} & $\boldsymbol{T_e,T_i}$ & $\boldsymbol{n_e}$, main ion $\boldsymbol{n_i}$ \\
\textbf{Mode} & evolving & predictive & predictive inside LCFS & predictive & predictive \\
\textbf{Name} & HFPS  & JETTO + NCLASS  & ESCO & TGLF-SAT2 & TGLF-SAT2 \\
\midrule
\textbf{Duty} & \textbf{Impurities} ($n_z$) & $\boldsymbol{V_\phi}$ & \textbf{Radiation} & \textbf{Heat sources} & \textbf{Particle sources} \\
\textbf{Mode} & predictive & n/a & predictive & predictive & predictive \\
\textbf{Name} & SANCO (N, W) evolved with TGLF-SAT2 & n/a & SANCO + ADAS & Ohmic, resistivity from NCLASS & FRANTIC (neutrals, energy 20 eV) + HPI2 (pellet, new parametrization) \\
\midrule
\textbf{Duty} & \textbf{Radial boundary condition} & \textbf{Time window} $\boldsymbol{\Delta t}$ & \textbf{MHD} & \textbf{Feedback controllers} \\
\textbf{Mode} & prescribed & $\Delta t = 5.5$\,s & prescribed trigger/predictive reconnection & $n_N$ and $n_W$ \\
\textbf{Name} & fixed $n_{e,\mathrm{sep}}$, $T_{e,\mathrm{sep}}$, $T_{i,\mathrm{sep}}$ & flat-top incl.\ pre- \& post-pellet phases & fix periodic trigger + Kadomtsev reconnection & Boundary influx tuned to fit core $T_e$ and $P_{rad}$ at fixed $v_{esc,\: boundary}$ \\
\bottomrule
\end{tabular*}
\end{table*}

In WEST, the coupling between impurities and the electron energy balance is particularly strong due to tungsten radiation. Especially in ohmic conditions, radiation can significantly constrain the achievable $T_e$ level and even modify profile shapes, motivating a consistent treatment of impurity transport and radiative losses in integrated modeling \cite{Maget_2022}. In the present predictive simulations, nitrogen and tungsten are evolved with SANCO and contribute self-consistently to the total radiated power.

Because the absolute impurity sources from wall interactions and seeding are not always directly constrained with sufficient accuracy, we adopt a pragmatic validation strategy: the impurity influx (or equivalently, the N and W content at the LCFS boundary condition) is adjusted within experimentally plausible bounds using feedback control. The scope is to reproduce consistently these experimentally measured quantities: both N and W concentrations are adjusted to match the measured $T_e (\rho_{tor}=0)$ level in the pre-pellet phase; as mentioned previously, the N boundary influx is then adjusted to match the experimentally reconstructed value of $Z_{eff}$ (since N is the most abundant impurity); and the W boundary influx is adjusted to match the experimentally reconstructed value of $P_{rad}$ (since W is the main radiator).

\subsection{Results}\label{subsec:hfps_results}
To obtain the final HFPS reference simulation, we followed a staged modeling strategy. We started from a set-up in which only the current diffusion was evolved predictively, while the remaining channels were kept interpretative. We then enabled predictive heat transport, followed by fully predictive heat and particle transport. Finally, impurity transport and radiation were included, yielding the complete self-consistent set-up discussed in the following.

To compare the simulated and experimental results we will use the Mean Absolute Percentage Error (MAPE) metric, denoted as $\mathrm{\sigma}$, which is defined as:
\begin{equation}
    \mathrm{\sigma} = \frac{1}{N}\sum_{i=1}^N \left| \frac{y_i - \hat{y}_i}{y_i} \right| \times 100\%
\end{equation}
where $y_i$ are the observed values, $\hat{y}_i$ are the predicted values, and $N$ is the total number of observations. This metric provides a percentage error between the predicted and observed values, allowing for an intuitive understanding of the accuracy of the predictions in relation to the actual measurements. When $\sigma_{global}$ appears, it means that the metric has been applied over an average of all time steps and all radial positions.
\subsubsection{Reference predictive simulation}\label{subsubsec:hfps_ref}
\begin{figure*}[!tbp]
\centering
\includegraphics[width=\linewidth]{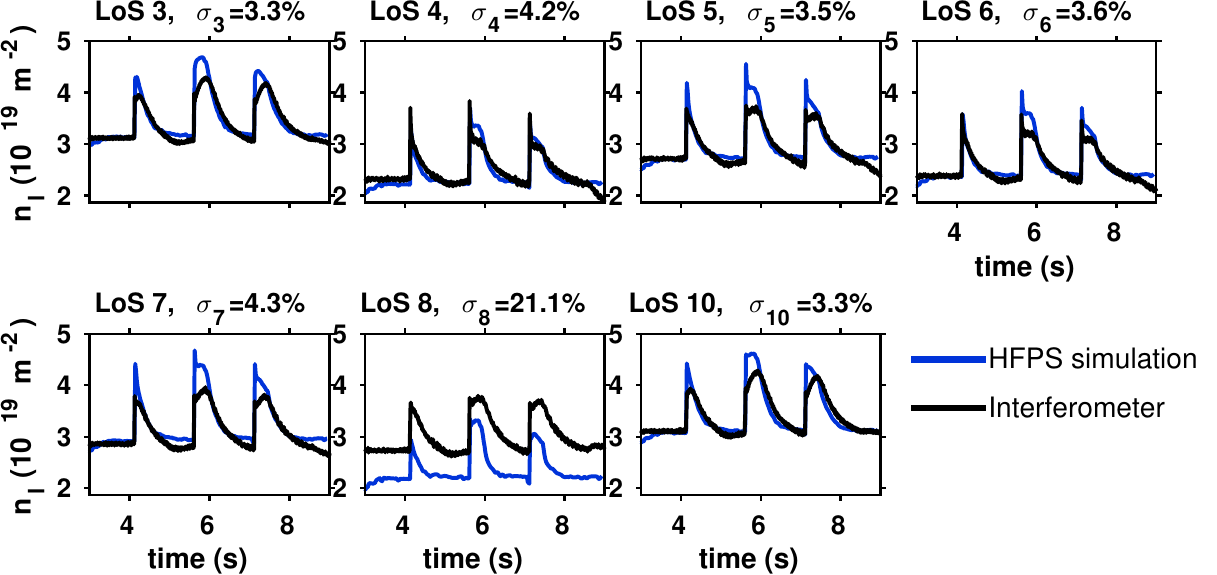}
\caption{Comparison of measured interferometry chords for LoS 3-8 and 10  (black lines) and synthetic line-integrated densities computed from the HFPS simulated $n_e(\rho,t)$ (blue lines). MAPE values ($\sigma_{LoS}$) are indicated for each line of sight throughout the whole time interval.}
\label{fig:hfps_interf}
\end{figure*}

Figure~\ref{fig:hfps_interf} compares the experimental interferometry chords with synthetic line-integrated densities computed from the HFPS-predicted $n_e(\rho,t)$. The synthetic diagnostic is obtained using the same procedure described in Section~\ref{sec:val} with the Syndi code \cite{Devynck2021}. The ability to reproduce multiple chords simultaneously indicates that the simulation captures the global density evolution associated with pellet fueling. The agreement is not uniform across all lines of sight: it is generally better for more core-weighted chords, and chords covering the upper part of the plasma (see Figure~\ref{fig:LoS}), therefore, it is notable the case of LoS~8, where the largest discrepancies are observed ($\sigma > 20\%$). However, for the rest of the chords, the agreement lies between $\sigma = 3.3\%$ and $\sigma = 4.3\%$, which is also notable. Lines of sight 1, 2 and 9, are excluded following the same criteria as in the stand-alone validation, since they are generally too edge-sensitive.

While the prompt rise and subsequent decay are well reproduced for the first pellet, the experimental signals exhibit a small post-pellet ``plateau'' following the second and third pellets (it can be appreciated in Figure~\ref{fig:hfps_interf} LoS 4, 5, 6 and 8), lasting approximately $300\,\mathrm{ms}$. In the present HFPS configuration, a transient change in edge fueling/recycling can only be represented through the boundary neutral particle source.  More specifically, the baseline simulation uses the prescribed separatrix conditions, so any transient increase in edge fueling needs to be represented explicitly through an extra boundary source term. We therefore mimicked this plateau by applying an additional neutral particle boundary influx contribution during the corresponding time window to describe the impact of enhanced recycling at the main wall.

From the perspective of the confined-plasma particle balance, this additional particle source is equivalent to an increased neutral influx at the edge and thus to enhanced recycling/edge fueling. Importantly, the plateau can be reproduced without modifying the pellet model parameters, which supports an interpretation of this feature as an edge-condition effect (recycling/pumping, peripheral neutral density, wall reservoir saturation) rather than as a deficiency of the pellet deposition model. In future HFPS studies, coupling to dedicated edge models such as EDGE2D-EIRENE \cite{Wiesen2006_EDGE2DEIRENE}, or to machine learning surrogates such as SOLPS-NN \cite{Wiesen_2024_AI}, could provide a more self-consistent description of recycling and separatrix boundary conditions, which are imposed as fixed in the present setup.

\begin{figure*}[!tbp]

  \centering
  \includegraphics[width=\linewidth]{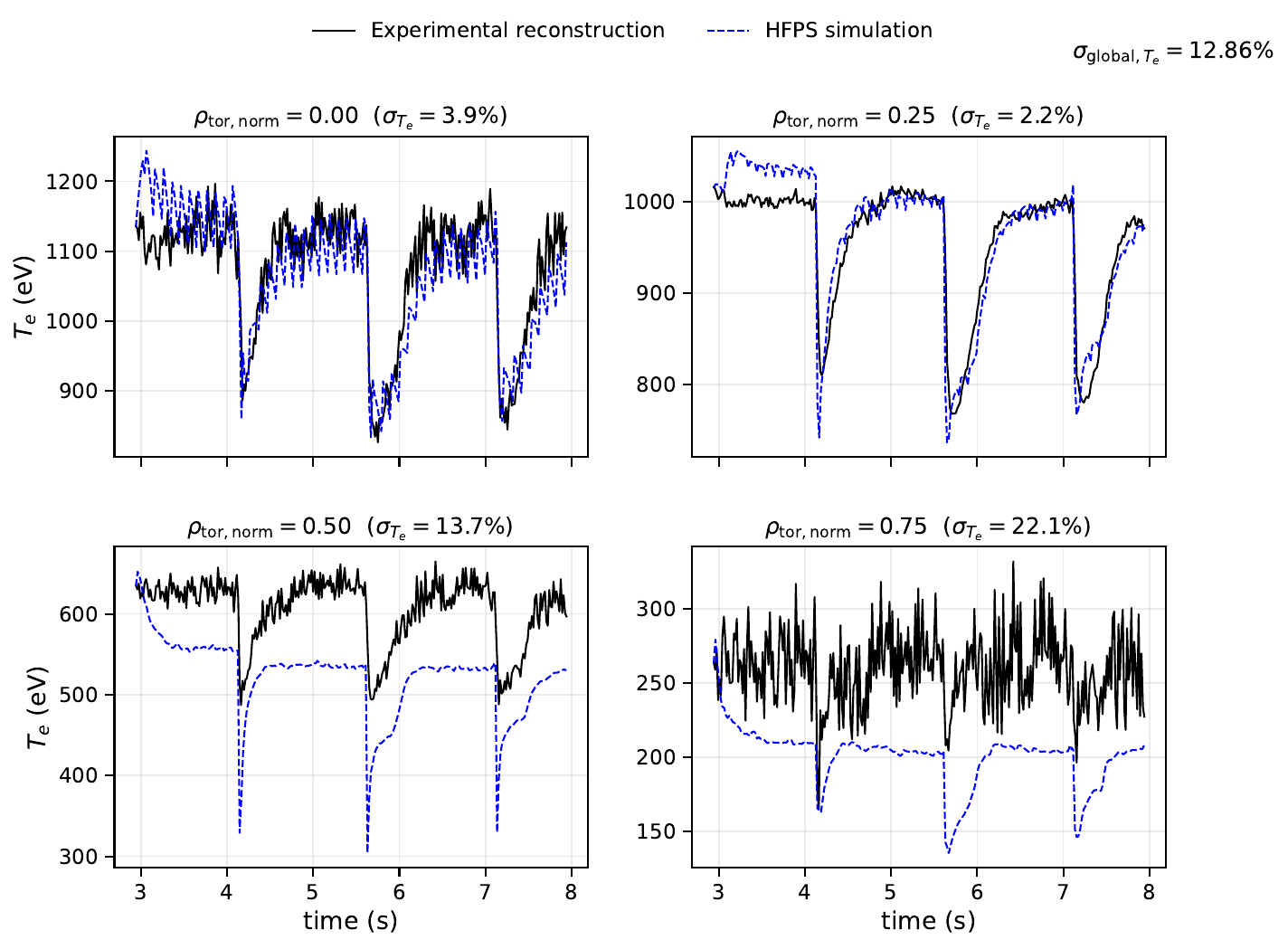}
  
  \caption{Time evolution of $T_e$ at $\rho_{tor, norm}=\{0, 0.25, 0.5, 0.75\}$, comparing experimental profile reconstruction (black) and HFPS predictive simulation (blue). MAPE values ($\sigma_{T_e}$) are indicated for each radius for the whole time interval, and as a global metric.}
  \label{fig:hfps_time_traces_Te}
\end{figure*}

Figure~\ref{fig:hfps_time_traces_Te} compares the simulated and measured time evolution of $T_e$ at selected radii, using the reconstructed experimental profiles as reference; the agreement clearly improves towards the core. The agreement is overall $\sigma_{global, T_e} = 12.86 \%$.

For $n_e$ and $T_e$, the accuracy is notable in the core ( $\rho_{tor, norm} \leq 0.5$ in Figure~\ref{fig:hfps_time_traces_Te}, and LoS 3, 5, 7, 10 in Figure~\ref{fig:hfps_interf}). Towards the edge ($\rho_{tor, norm} > 0.5$ in Figure~\ref{fig:hfps_time_traces_Te}, and LoS 4, 6, 8 in Figure~\ref{fig:hfps_interf}), the discrepancies increase, specially in the case of $T_e$ with $\rho_{tor, norm} > 0.5$ and the interferometry LoS 8. In the case of $T_e$, the discrepancies are consistent with the use of fixed separatrix boundary conditions that constrain the edge evolution and can limit the ability to match the experimental dynamics locally; also, they could be linked to an overestimation on the tungsten boundary influx, causing an underprediction in $T_e$ at the edge. Regarding the offset observed for LoS 8 in Figure~\ref{fig:hfps_interf}, as mentioned in Section~\ref{sec:val_results}, this could be due to an equilibrium reconstruction mismatch, rather than an edge-related effect, specifically an underestimation of the elongation of the lower part of the plasma.

Overall, the predictive simulation captures (i) the prompt density rise associated with the pellet deposition, (ii) the subsequent redistribution timescale, and (iii) the correlated $T_e$ transient and recovery.

\begin{figure*}[!tbp]

  \centering
  \includegraphics[width=\linewidth]{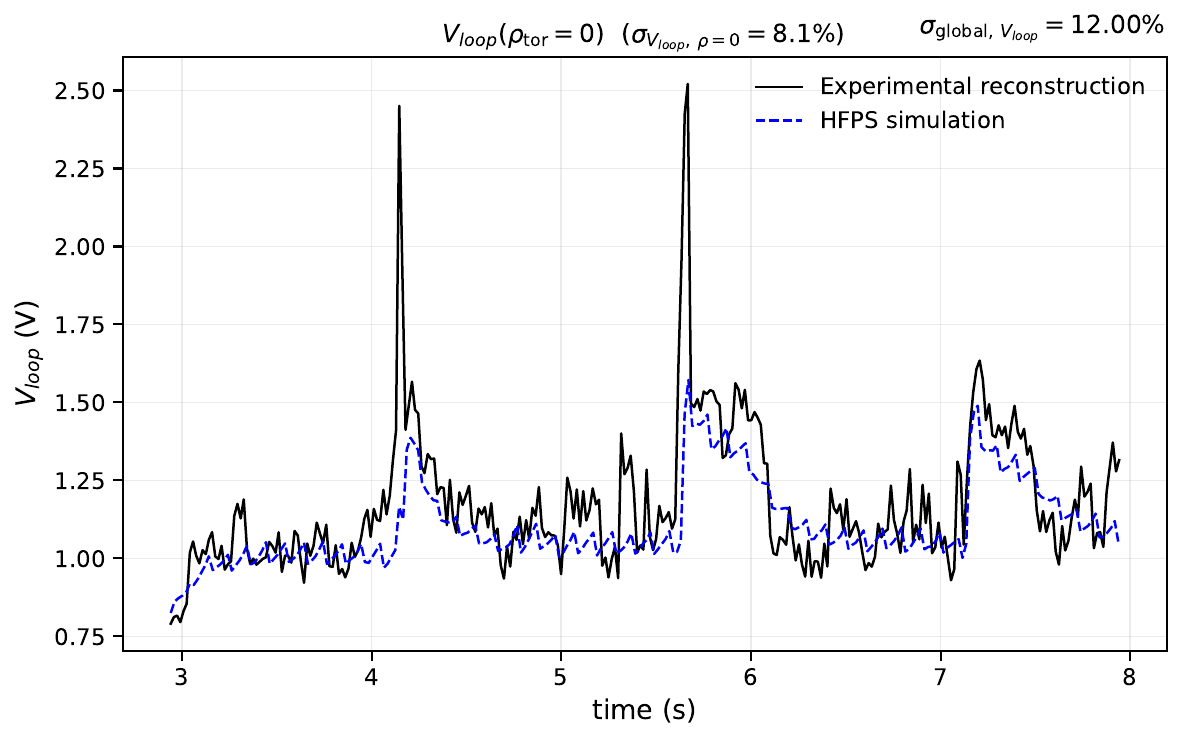}
  
  \caption{Time evolution of $V_{loop}$ at $\rho=0$, comparing experimental profile reconstruction (black) and HFPS predictive simulation (blue). MAPE values ($\sigma_{V_{\mathrm{loop}}}$) are indicated for $\rho_{tor, norm}=0$ throughout the whole time interval, and as a global metric.}
  \label{fig:hfps_time_traces_Vloop}
\end{figure*}

Additional support for the consistency of the integrated simulation is provided by the agreement on global power-balance quantities. Figure~\ref{fig:hfps_time_traces_Vloop} shows that the predicted loop voltage at core ($\rho_{tor, norm} = 0$) follows the experimental evolution with reasonable agreement ( $\sigma_{global,V_{\mathrm{loop}}} = 12.0 \%$ ). $V_{\mathrm{loop}}$ is directly related to the inductive electric field and therefore to the ohmic power balance, so this agreement indicates that the simulated current-diffusion and resistive-heating dynamics remain consistent with the experimental discharge evolution.

\begin{figure*}[!tbp]

  \centering
  \includegraphics[width=\linewidth]{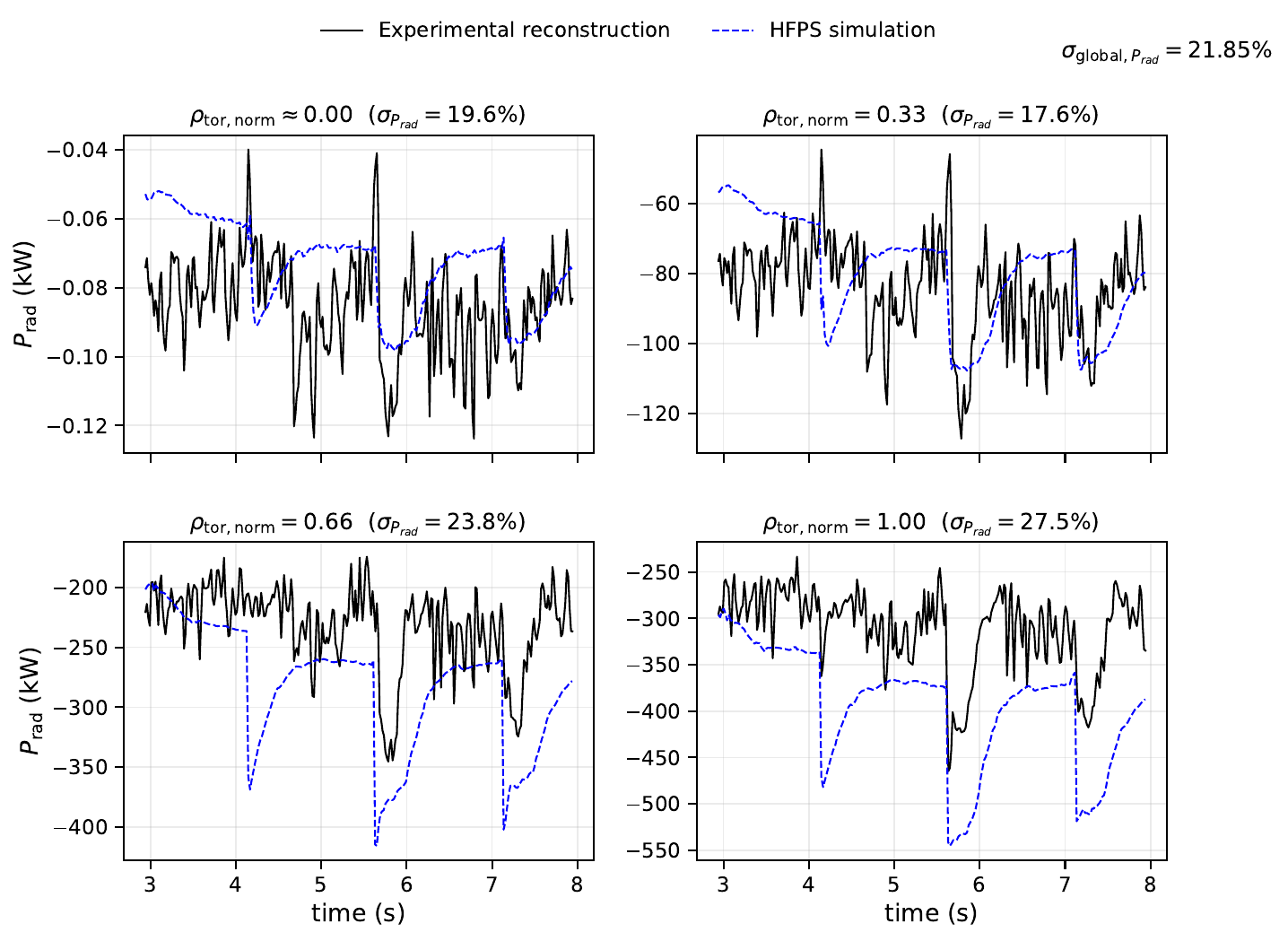}
  
\caption{Time evolution of $P_\mathrm{rad}$ at $\rho=\{0.0, 0.33, 0.66, 1\}$, comparing experimental profile reconstruction (black) and HFPS predictive simulation (blue).}
\label{fig:hfps_time_traces_Prad}
\end{figure*}

Likewise, Figure~\ref{fig:hfps_time_traces_Prad} shows the reproduction of the radiated power time evolution (reconstructed out of measurements from the WEST bolometer \cite{Devynck2021}). The accuracy is lower than for the other quantities, but still reproduces the order of magnitude, specially closer to the core. This agreement is closely connected to the core electron temperature, since impurity cooling rates strongly depend on Te and excessive radiative losses tend to lower $T_e$; an accurate description of this power balance is therefore essential. As commented in section~\ref{subsec:hfps_setup}, since tungsten is the dominant radiator in this WEST ohmic plasma, the agreement on $P_{\mathrm{rad}}$ also supports the tungsten concentration modeled in the simulation. It is worth pointing out that the tungsten boundary influx in the simulation might have been slightly overestimated, since the absolute values of $P_rad$ in the edge are overpredicting, while $T_e$ is underpredicting in the edge too.

These results support that the combination of the HPI2 pellet particle source and TGLF-SAT2 turbulent transport captures the main multi-channel response of this ohmic, pellet-fueled plasma.

\subsubsection{Discussion: complementarity of HPI2 and TGLF-SAT2 in an ohmic pellet-fueled plasma}\label{subsubsec:hfps_discussion}

This exercise provides a coupled validation of (i) a physics-based pellet deposition model and (ii) a multi-channel turbulent transport model in an ohmic regime of direct relevance for reactor-oriented scenarios, where pellet fueling is expected to dominate over direct neutral penetration. Rather than attributing the agreement to a single ingredient, the simultaneous agreement obtained on local kinetic profiles and interferometry signals supports the following points.

First, the HPI2 deposition localization and ablation timescale are compatible with the experimentally inferred prompt density rise when embedded in an integrated modeling environment. This extends the stand-alone validation by demonstrating that the pellet source remains consistent once the subsequent redistribution is governed by self-consistent turbulent transport.

Second, the results indicate that TGLF-SAT2 captures the dominant transport response in this WEST ohmic regime, in particular the relaxation of pellet-induced density gradients and the associated timescales. Similar integrated modeling studies have shown that coupling a pellet deposition model to a quasilinear turbulent transport model can reproduce the measured density evolution over pellet cycles, and that synthetic-diagnostic validation against interferometry provides a robust route to assess the coupled source-transport system \cite{Marin_2021}.

Third, the analysis highlights the importance of separating source-related and edge-condition effects. While the synthetic interferometry chords reproduce the ablation-phase density increment consistently across pellets, the post-pellet plateau observed after the second and third pellets required an additional edge particle source in the simulation. Since this feature can be recovered without modifying HPI2 parameters, it supports an interpretation in terms of evolving edge recycling/pumping conditions rather than deficiencies of the deposition model. This illustrates the benefit of validating the pellet deposition primarily on the ablation-phase density rise, and using the post-ablation evolution to constrain edge and transport modeling assumptions.

Finally, the role of impurities is essential in a W environment: obtaining the correct $T_e$ level requires a consistent radiation balance. Here this is addressed by evolving N and W with SANCO while constraining their net level within experimentally plausible bounds. This strategy keeps the validation focus on the coupled response of fueling and turbulent transport, while retaining the key physics channel through which impurities regulate the ohmic temperature evolution.

The achievable validation accuracy is limited by experimental uncertainties in (i) reconstructed kinetic profiles ($T_e$, $n_e$, and $T_i$), (ii) equilibrium reconstruction and mapping to $\rho$, and (iii) pellet mass/velocity estimates, as well as by modeling assumptions at the boundary (fixed separatrix conditions and simplified neutral treatment).

\section{Conclusions and outlook}\label{sec:conc}

This work addresses a central challenge for the exploitation of ITER class plasmas: predicting, in a consistent and computationally viable manner, the particle source associated with pellet injection and its coupling to turbulent transport, impurity evolution and the power balance. Within this framework, the main methodological outcome is an improvement of the physics-based pellet code HPI2, in which the plasmoid release spatial step is determined self-consistently from ablation physics through
\begin{equation}
dx_{var} = v_{\mathrm{pel}}\, t_{\mathrm{exit}},
\end{equation}
optionally including a scaling factor to trade accuracy for computational cost. This modification removes an ad-hoc numerical parameter, improves model robustness and strengthens extrapolability when HPI2 is employed within predictive integrated modeling frameworks.

Validation has been carried out at two levels against the upper high field side pellet-fueled ohmic WEST discharge \#58656. In stand-alone mode, HPI2 reproduces the pellet-induced line-integrated density increments measured by interferometry using synthetic diagnostics within an average of $\sim 10\%$ error, and enables key pellet parameters (mass and velocity) to be constrained within experimental uncertainties. When HPI2 is coupled within the High Fidelity Pulse Simulator (HFPS) workflow (IMAS/JINTRAC), using HFPS as the 1.5D transport solver, TGLF-SAT2 for turbulent transport, and SANCO for impurity transport and radiation, the simulations simultaneously capture (i) the density rise associated with pellet deposition, (ii) the characteristic relaxation/redistribution timescale, and (iii) the correlated $T_e$ transient. Sufficiently accurate simulation results are obtained with reconstructed $T_e$ profiles ($\sigma_{T_e}=12.86\%$), with line integrated density measurements from multiple interferometry chords ($\sigma_{n_l}\approx 4\%$ for all studied lines of sight except 8 $\sigma_{n_l, 8}\approx 21\%$), with reconstructed loop voltage ($\sigma_{V_{loop}}=12.0\%$), and with bolometry reconstructed radiative power ($\sigma_{P_{rad}}=21.85\%$). Overall, the results support that the upgraded HPI2 particle source remains consistent when embedded in a predictive integrated modeling environment, even in a W-wall scenario where impurity radiation strongly constrains the thermal evolution. In addition, the analysis highlights the importance of edge conditions for the post-pellet evolution: a short density plateau observed experimentally after the second and third pellets can be reproduced in HFPS by introducing an additional edge particle source (gas fueling) in the corresponding time window, without modifying the HPI2 deposition model parameters. This supports an interpretation of the plateau in terms of evolving recycling/pumping conditions, rather than as a deficiency of the pellet deposition calculation. It also indicates that future predictive studies would benefit from coupling HFPS to more advanced edge/SOL models, to treat recycling, and specially separatrix boundary conditions self-consistently instead of prescribing them as fixed inputs.

From a fusion perspective, this kind of end-to-end validation is a necessary step to reduce the risk associated with extrapolating pellet-based density control to larger and higher-power devices, where operational margins are tighter and the interplay between sources, transport and radiation becomes more critical. A more robust pellet deposition model, less dependent on numerical tuning, facilitates systematic scenario studies, actuator design (pellet frequency, mass and velocity) and, in the medium term, the development of integrated control strategies combining pellet fueling with the rest of plasma actuators.

\paragraph{Outlook.}

As a natural continuation of this work, (i) the sensitivity of the HPI2-HFPS coupling to separatrix boundary conditions and to more realistic neutrals/edge modeling could be explored, with the aim of improving agreement in the edge plasma and outer part of the core plasma ; (ii) validation could be extended to phases with auxiliary heating and to alternative injection geometries (including LFS and other configurations) to cover a wider range of conditions; (iii) more reliable pellet velocity estimates for further studies; and (iv) the transferability of this validation could be assessed in ITER and other larger-device scenario and control studies, where reliable predictive modeling of pellet fueling will be critical for sustained operation and performance optimization.

\section*{Acknowledgements}
The authors acknowledge the valuable advice and support provided by colleagues at CEA Cadarache, the ITER Organization, and UKAEA. The authors also thank the members of the EUROfusion TSVV11 group for fruitful discussions and continuous technical input throughout this work. The authors express special thanks to Patrick Maget (CEA Cadarache) and Francis Casson (UKAEA) for their valuable advice and support during this study.

This work has been carried out within the framework of the EUROfusion Consortium, funded by the European Union via the Euratom Research and Training Programme (Grant Agreement No 101052200 — EUROfusion). Views and opinions expressed are however those of the author(s) only and do not necessarily reflect those of the European Union or the European Commission. Neither the European Union nor the European Commission can be held responsible for them.

The views and opinions expressed herein do not necessarily reflect those of the ITER Organization
% \printbibliography
\bibliographystyle{iopart-num}
\bibliography{bibliography}

\newpage

\end{document}